\documentclass[reprint,pra,notitlepage,superscriptaddress]{revtex4-1}

\usepackage{booktabs} 
\usepackage{dcolumn}
\usepackage{bm}

\usepackage[normalem]{ulem}

\usepackage{colonequals}
\usepackage[colorlinks]{hyperref}
\usepackage{hypcap}
\usepackage{enumitem}

\usepackage{amsmath}
\usepackage{amsfonts,amssymb,amsthm, bbm,braket}
\usepackage{graphicx}   
\usepackage{amsbsy} 
\usepackage{accents}
\usepackage[bold]{hhtensor}  

\usepackage{algpseudocode}

    \newcommand{\inlinecomment}[1]{\Comment {\footnotesize #1} \normalsize}

\usepackage{multirow}
\usepackage{subcaption}
\DeclareCaptionType{algorithm}
\DeclareCaptionSubType{algorithm}

\setlist[enumerate]{itemsep=0mm}

\newcommand{\ketbra}[2]{\left|#1\right\rangle\!\left\langle #2\right|}

\newcommand{\suppress}[1]{}

\def\squareforqed{\hbox{\rlap{$\sqcap$}$\sqcup$}}
\def\qed{\ifmmode\squareforqed\else{\unskip\nobreak\hfil
\penalty50\hskip1em\null\nobreak\hfil\squareforqed
\parfillskip=0pt\finalhyphendemerits=0\endgraf}\fi}

\newcommand{\eq}[1]{\hyperref[eq:#1]{(\ref*{eq:#1})}}
\renewcommand{\sec}[1]{\hyperref[sec:#1]{Section~\ref*{sec:#1}}}
\newcommand{\app}[1]{\hyperref[app:#1]{Appendix~\ref*{app:#1}}}
\newcommand{\tab}[1]{\hyperref[tab:#1]{Table~\ref*{tab:#1}}}
\newcommand{\fig}[1]{\hyperref[fig:#1]{Figure~\ref*{fig:#1}}}
\newcommand{\thm}[1]{\hyperref[thm:#1]{Theorem~\ref*{thm:#1}}}
\newcommand{\lem}[1]{\hyperref[lem:#1]{Lemma~\ref*{lem:#1}}}
\newcommand{\cor}[1]{\hyperref[cor:#1]{Corollary~\ref*{cor:#1}}}
\newcommand{\defn}[1]{\hyperref[def:#1]{Definition~\ref*{def:#1}}}
\newcommand{\alg}[1]{\hyperref[alg:#1]{Algorithm~\ref*{alg:#1}}}

\newcommand{\T}{\mathrm{T}}

\renewcommand{\H}{\mathcal{H}}
\newcommand{\Lin}{\mathcal{L}}

\renewcommand{\eprint}[1]{\href{http://arxiv.org/abs/#1}{#1}}

\newcommand{\Tr}{\operatorname{Tr}}

\usepackage[bold]{hhtensor}
\usepackage{braket}
\input{Qcircuit}

\newcommand{\swapgt}{\textsc{swap}}

\newcommand{\dd}{\mathrm{d}}
\newcommand{\ad}{\operatorname{ad}}

\def\Sket#1{\left|#1\right\rrangle}
\makeatletter
\def\Decl@Mn@Delim#1#2#3#4{%
  \if\relax\noexpand#1%
    \let#1\undefined
  \fi
  \DeclareMathDelimiter{#1}{#2}{#3}{#4}{#3}{#4}}
\def\Decl@Mn@Open#1#2#3{\Decl@Mn@Delim{#1}{\mathopen}{#2}{#3}}
\def\Decl@Mn@Close#1#2#3{\Decl@Mn@Delim{#1}{\mathclose}{#2}{#3}}
\DeclareFontFamily{OMX}{MnSymbolE}{}
\DeclareFontShape{OMX}{MnSymbolE}{m}{n}{
    <-6>  MnSymbolE5
   <6-7>  MnSymbolE6
   <7-8>  MnSymbolE7
   <8-9>  MnSymbolE8
   <9-10> MnSymbolE9
  <10-12> MnSymbolE10
  <12->   MnSymbolE12}{}
\DeclareFontShape{OMX}{MnSymbolE}{b}{n}{
    <-6>  MnSymbolE-Bold5
   <6-7>  MnSymbolE-Bold6
   <7-8>  MnSymbolE-Bold7
   <8-9>  MnSymbolE-Bold8
   <9-10> MnSymbolE-Bold9
  <10-12> MnSymbolE-Bold10
  <12->   MnSymbolE-Bold12}{}
\DeclareSymbolFont{mnsymbols}  {OMX}{MnSymbolE}{m}{n}
\Decl@Mn@Open {\lsem}               {mnsymbols}{'102}
\Decl@Mn@Close{\rsem}               {mnsymbols}{'107}
\Decl@Mn@Open {\llangle}            {mnsymbols}{'164}
\Decl@Mn@Close{\rrangle}            {mnsymbols}{'171}
\makeatother

\newlength{\dhatheight}
\newcommand{\hhatDS}[1]{%
    \settoheight{\dhatheight}{\ensuremath{\hat{#1}}}%
    \addtolength{\dhatheight}{-0.25ex}%
    \hat{\vphantom{\rule{1pt}{\dhatheight}}%
    \smash{\hat{#1}}}}
\newcommand{\hhatTS}[1]{%
    \settoheight{\dhatheight}{\ensuremath{\hat{#1}}}%
    \addtolength{\dhatheight}{-0.25ex}%
    \hat{\vphantom{\rule{1pt}{\dhatheight}}%
    \smash{\hat{#1}}}}    
\newcommand{\hhatS}[1]{%
    \settoheight{\dhatheight}{\ensuremath{\scriptstyle{\hat{#1}}}}%
    \addtolength{\dhatheight}{-0.175ex}%
    \hat{\vphantom{\rule{1pt}{\dhatheight}}%
    \smash{\hat{#1}}}}
\newcommand{\hhatSS}[1]{%
    \settoheight{\dhatheight}{\ensuremath{\scriptscriptstyle{\hat{#1}}}}%
    \addtolength{\dhatheight}{-0.07ex}%
    \hat{\vphantom{\rule{1pt}{\dhatheight}}%
    \smash{\hat{#1}}}}
\newcommand{\hhat}[1]{\mathchoice{\hhatDS{#1}}{\hhatTS{#1}}{\hhatS{#1}}{\hhatSS{#1}}}

\global\def \arxivmode {}

\ifx \arxivmode \undefined
\else
  
  \newcommand\arxivonly[1]{#1}
  \newcommand\prlonly[1]{}
\fi

\ifx \prlmode \undefined
\else
  
  \newcommand\arxivonly[1]{}
  \newcommand\prlonly[1]{#1}
  \renewcommand{\section}[1]{}
\fi

\begin{document}

\title{Quantum Hamiltonian Learning Using Imperfect Quantum Resources}
\author{Nathan Wiebe}
\affiliation{Quantum Architectures and Computation Group, Microsoft Research, Redmond, WA 98052, USA}
\affiliation{Department of Combinatorics \& Optimization, University of Waterloo, Ontario N2L 3G1, Canada}
\affiliation{Institute for Quantum Computing, University of Waterloo, Ontario N2L 3G1, Canada}
\author{Christopher Granade}
\affiliation{Department of Physics, University of Waterloo, Ontario N2L 3G1, Canada}
\affiliation{Institute for Quantum Computing, University of Waterloo, Ontario N2L 3G1, Canada}
\author{Christopher Ferrie}
\affiliation{
Center for Quantum Information and Control,
University of New Mexico,
Albuquerque, New Mexico, 87131-0001}
\author{David Cory}
\affiliation{Department of Chemistry, University of Waterloo, Ontario N2L 3G1, Canada}
\affiliation{Institute for Quantum Computing, University of Waterloo, Ontario N2L 3G1, Canada}
\affiliation{Perimeter Institute, University of Waterloo, Ontario N2L 2Y5, Canada}

\begin{abstract}
Identifying an accurate model for the dynamics of a quantum system is a vexing problem that underlies a range of problems in experimental physics and quantum information theory.
Recently, a method called \emph{quantum Hamiltonian learning} has been proposed by the present authors that uses quantum simulation as a resource for modeling an unknown quantum system.
This approach can, under certain circumstances, allow such models to be efficiently identified.
A major caveat of that work is {the assumption of that all elements of the protocol are noise--free}.
Here, we show that quantum Hamiltonian learning can tolerate substantial amounts of depolarizing noise and show numerical evidence that it can tolerate noise drawn from other realistic models. 
We further provide evidence that the learning algorithm will find a model that is maximally close to the true model in cases where the hypothetical model lacks terms present in the true model.  
Finally, we also provide numerical evidence that the algorithm works for non--commuting models.  This work illustrates that quantum Hamiltonian learning can be performed using realistic resources and suggests that even imperfect quantum resources may be valuable for characterizing quantum systems.
\end{abstract}
\maketitle

\section{Introduction}
The challenges faced by experimentalists trying to learn an appropriate Hamiltonian model for a large quantum system can be quite daunting.  Traditional techniques, such as tomography, rapidly become infeasible as the number of qubits increases.  To make matters worse, the dynamics of such quantum systems cannot even be simulated efficiently using existing methods without making strong assumptions.
This raises an important question: how can you decide the properties of a model that is too complex to even simulate?

This is not simply a point of theoretical interest.  Present day experiments are already operating in regimes that are challenging for classical supercomputers to simulate~\cite{bollinger2013quantum,britton2012engineered}, and near future experiments will soon be well beyond their capabilities~\cite{korenblit2012quantum,PhysRevA.87.013422,islam2013emergence,tillmann2013experimental}.  The stakes are further compounded by the fact that many of these systems are intended as quantum simulators or demonstrations of the supremacy of quantum information processing over classical information processing, and none of these demonstrations can be considered compelling unless their predictions can be independently verified~\cite{hauke_trust_2012,gogolin_boson_2013,ududec_equilibration_2013}.  A resolution to this problem is thus urgently needed.

A natural solution to this problem is to leverage the inherent power of quantum systems as a resource to characterize other quantum systems.
The idea behind this approach is simple: in order to learn a Hamiltonian model, you build a trustworthy quantum simulator for that class of models and use it to make dynamical predictions about a hypothetical model for the quantum system.  This quantum simulator need not necessarily be a quantum computer,  but it must be ``trusted'' that the dynamical map that it implements is sufficiently close to that of the ideal model.

In~\cite{QHL}, we provided a concrete way of implementing this procedure using Bayesian inference, wherein trusted quantum simulators are used to compute the probability that a hypothetical model would yield the observed measurement outcome.  The approach is shown to be remarkably efficient at learning Hamiltonian parameters and resilient to some forms of noise, such as shot noise in the computation of the probabilities.  Two implementations of quantum Hamiltonian learning are proposed in~\cite{QHL}.  The first involves simply using a quantum simulator as a resource for computing likelihood functions that appear in Bayesian inference, or more concretely, the Hamiltonian inference protocol in~\cite{granade_robust_2012}.  The second approach goes beyond this, by allowing the quantum simulator to interact with the experimental system via a \swapgt~gate.  This approach is called \emph{interactive} quantum Hamiltonian learning, and it is shown to be more stable and efficient than its non--interactive brethren.  Here, we go beyond these results and show that interactive quantum Hamiltonian learning is resilient to realistic sources of noise that can arise in the protocol.
This not only illustrates that interactive experiments can be performed with realistic quantum resources but also suggests that they could be performed with existing or near future quantum systems.

Before ending this introduction with an outline of the paper, we briefly comment on the relation to other learning methods which seek to reduce the cost of characterization and validation.  These include identifying stabilizer states \cite{Gottesman2008Identifying,daSilva2011Practical}; tomography for matrix product states \cite{Cramer2010Efficient}; tomography for permutationally invariant states \cite{Toth2010Permutationally}; learning local Hamiltonians \cite{daSilva2011Practical};
tomography for low-rank states via compressed sensing \cite{Flammia2012Quantum}; and
tomography for multi-scale entangled states
\cite{LandonCardinal2012Practical}.  Several of these methods use quantum resources to accellerate quantum the characterization such as the matrix product state tomography method of~\cite{Cramer2010Efficient} and the direct quantum process tomography method of~\cite{mohseni2006direct}.  Direct quantum process tomography is in some senses analogous to our work because it uses two--qubit (or qudit) interactions to infer the dynamics of an unknown system thereby removing the need to perform a computationally expensive inversion procedure.

The key difference between prior works and ours is that the above techniques employ efficient \emph{classical} simulation algorithms which propagate efficient representations of the state vector to calculate of the probabilities defining the likelihood function.  Whereas, we evaluate the likelihood function using \emph{quantum} resources.  

We layout the paper as follows.  We discuss the theory of Bayesian inference our quantum Hamiltonian learning scheme (as well as the prior classical result in~\cite{granade_robust_2012}) uses in~\sec{cle}.  \sec{qhl} reviews the results in~\cite{QHL}, which provides a method for making Bayesian inference practical by using quantum simulation to evaluate the likelihood function.  We also provide an explicit algorithm for the procedure in~\sec{qhl}.  We then present  numerical evidence in~\sec{nonoise} that shows the quantum Hamiltonian learning algorithm (QHL) can rapidly infer a model for Ising and transverse Ising model Hamiltonians.  \sec{noise} provides theoretical evidence that the learning algorithm is robust to depolarizing noise and that realistic noise models for the \swapgt~gates used in interactive quantum Hamiltonian learning experiments do not prevent the algorithm from learning at an exponential rate.
{Finally, in \sec{model-selection}, we consider the performance of QHL when the model
being used differs from the physics governing an experiment.}

\section{Classical Hamiltonian Learning}\label{sec:cle}
Many approaches for learning the Hamiltonian of an uncharacterized quantum system have been considered, but only recently have ideas and methods from statistical inference been applied to this problem despite their ubiquity in machine learning and related fields~\cite{hentschel_machine_2010,hentschel_efficient_2011,sergeevich_characterization_2011,lovett_differential_2013,granade_robust_2012}.
Here, we consider a Bayesian approach to parameter estimation.  This approach, in essence, uses Bayes' theorem to give the probability that a hypothesis about the Hamiltonian $H$ is true, given the evidence that has been accumulated through experimentation.  Or, in other words, it provides an approximation to the Hamiltonian (from a class of Hamiltonians) that is most likely to yield the observed experimental data. 

Bayesian inference problems are specified by two quantities:
\begin{enumerate}
\item A \emph{prior} probability distribution $\Pr(H)$ that encodes the a priori confidence that a given Hamiltonian model $H$ is correct. This is chosen by the experimenter, but can always be taken to be a least informative (typically uniform) distribution utilizing no prior assumptions. 
\item A \emph{likelihood function} $\Pr(D|H)$ that returns the probability that outcome $D$ is observed given $H$ is the true Hamiltonian.  This is not chosen; it is prescribed by quantum mechanics via the Born rule.
\end{enumerate}

Bayes' theorem states that the probability that a given hypothetical Hamiltonian $H$ is correct given the observed data $D$ can be computed from the prior and the likelihood function via
\begin{equation}
\Pr(H|D) =\frac{\Pr(D|H)\Pr(H)}{\Pr(D)}=\frac{\Pr(D|H)\Pr(H)}{\int \Pr(D|H)\Pr(H)\,\mathrm{d} H}.\label{eq:bayes}
\end{equation}
The inference process then involves repeating the above process for each observed datum after setting $\Pr(H) \gets \Pr(H|D)$, where each such repetition is known as an update.
The probability distribution will typically converge to a sharply peaked distribution about the true model parameters (unless uninformative experiments are chosen, or the learning problem is degenerate) as the number of updates increases.  This procedure has been shown to be extremely effective at Hamiltonian learning: only a few hundred experiments can lead to an accurate estimates of Hamiltonian.  In contrast, traditional methods can require billions of measurement outcomes to achieve comparable precision~\cite{granade_robust_2012,QHL}.

An important feature to note is that Bayesian inference reduces the problem of inference to a problem in simulation.  In cases where the likelihood can be easily computed, Bayesian inference will often be well suited for the problem; whereas it is ill--suited when the likelihood function is intractable.  This is the key insight behind the entire quantum Hamiltonian learning approach.

A typical Hamiltonian inference problem involves
evolving a known initial state $\ket{\psi}$ under an unknown Hamiltonian $H$, then measuring against
a fixed basis $\{\ket{D}\}$. The Hamiltonian is then inferred from the measurement statistics.  In Bayesian Hamiltonian inference, the likelihood function for such experiments is given by the Born rule as
as
\begin{equation}
\Pr(D|H)=|\bra{D}e^{-iHt}\ket{\psi}|^2.
\end{equation}

The final issue that needs to be considered is that the probability distributions must be discretized in order to make Bayesian updating tractable.  We use a finite particle approximation to the probability distributions known as the Sequential Monte Carlo approximation (SMC), in which we draw samples from the initial prior distribution.
Each such particle $i$ drawn from the initial prior is assigned a weight $w_i = 1 / N$,
and is then updated using Bayes' rule as data is collected.

A common problem with these methods is that the effective sample size $N_{\text{ess}} = \sum_i 1 / {w_i^2}$
of the approximation becomes small as data is incorporated, such that the approximation
becomes \emph{impoverished}.
To rectify this and to recover numerical stability, we employ a resampling method proposed by Liu and West that changes the particle positions in order to concentrate particles in regions of high posterior probability by interpreting the posterior at any given
time step as a mixture of the SMC-approximated posterior and a multivariate normal having the same mean and covariance
\cite{west_approximating_1993,Liu2000Combined}. The quantity $a$ allows the resampling algorithm to smoothly interpolate between sampling from the original SMC distribution and a Gaussian distribution; as $a \to 1$, the resampling algorithm draws particles from the SMC approximation to the posterior, whereas the resampling algorithm draws particles from a normal distribution with the same mean and variance as the posterior distribution in the limit $a \to 0$.
Here, we find that the approximate normality of the problem allows for us to take $a = 0.9$ such that we need fewer particles
for each simulated run of our algorithm.
By drawing new particles from the resampling mixture distribution,
the first two moments of the posterior are manifestly preserved, but the particle approximation is \emph{refreshed}
such that the effective sample size of the particle approximation is increased.

This resampling procedure is essential because, with high probability, none of the discrete particles used to represent the probability distribution will coincide with the correct model parameters.  This means that as time progresses, the algorithm will become more sure that each of the particles does not hold the correct model parameters but will not be able to accurately estimate the true parameters. Thus, the refresh of the effective sample size allowed by the resampling step
avoids this problem by adaptively changing the discretization to reflect the tighter posterior distributions afforded
by experimental knowledge.

Further discussion of the technical details of these methods can be found in~\cite{granade_robust_2012,Liu2000Combined}.  After this discretization process is done, Bayesian updating (and experiment design~\cite{granade_robust_2012}) can be applied to learn the Hamiltonian.

As an example of how the inference procedure may be used in practice, imagine that an experimentalist suspects that their system is an Ising model with Hamiltonian
\begin{equation}
H=\sum_{i=1}^{n-1} \frac{\pi J_{i}}{2} \sigma_z^{(i)}\sigma_z^{(i+1)}.\label{eq:isingline}
\end{equation}
Bayesian inference can then be used to find the most likely parameters $\{J_i\}$ given the experimental data and any prior knowledge about these parameters. Here and in the majority of the numerical cases that we consider, we take the pessimistic assumption that the experimentalist is maximally ignorant of the Hamiltonian given some
physical constraints and reflects this knowledge by choosing $\Pr(H)$ to be the uniform distribution
over all Hamiltonian operators obeying these constraints.

In general, we will denote the parameters for a Hamiltonian to be a vector $\vec{x}\in \mathbb{R}^d$ and that the corresponding Hamiltonian is $H(\vec{x})$.  For example, in this example $d=n-1$, $\vec{x} =[J_1,\ldots, J_{n-1}]$ and $H(\vec{x}) = \vec{x} \cdot [\sigma_z^{(1)}\sigma_z^{(2)},\ldots,\sigma_z^{(n-1)}\sigma_z^{(n)}]$.  An appropriate choice for the initial state is $\ket{+}^{\otimes n}$, and similarly computational basis measurements are suitable because these state preparations and measurements lead to different Hamiltonian parameterizations resulting in different measurement outcomes.  In contrast, computational basis state preparations are innapropriate because the Hamiltonian has a trivial action on them.

Bayesian inference is well suited for learning a concise parameterization of a Hamiltonian within a family of potential models.  In addition, region estimates for the true Hamiltonian can be easily found from the posterior distribution, which allows the confidence in the final inferred Hamiltonian \cite{ferrie_hpdregions_2013,granade_robust_2012} to be quantified.  In particular, we advocate for a concise representation of our uncertainty through the ellipse defined by the posterior covariance \cite{granade_robust_2012}, which was shown to be nearly optimal in terms of capturing the densest region of posterior probability \cite{ferrie_hpdregions_2013}, such that covariance ellipsoids provide a very good approximation to the highest-power credible region estimators. Moreover, by the use of a clustering algorithm, this can be extended to allow for efficient region estimation over multimodal distibutions. Thus, in addition to providing a fast method for inferring the form of the Hamiltonian, Bayesian inference also naturally gives an estimate of the uncertainty of the result, unlike most tomographic approaches. 

In practice, the posterior distribution tends to converge to a unimodal distribution that is, to a good approximation, Gaussian.  Under this assumption, an error ellipsoid that contains a ratio ${\rm erf}(Z/\sqrt{2})^d$ of the total aposteri probability is given by the set of all $\vec{x}$ that obey~\cite{granade_robust_2012}
\begin{equation}
(\vec{x}-\vec{\mu})^T\Sigma^{-1}(\vec{x}-\vec{\mu})\le Z^2,
\end{equation}
where $\vec{\mu}$ is the posterior mean and $\Sigma$ is the posterior covariance matrix.  Computation of the posterior mean and the inverse of the covariance matrix is efficient for fixed particle number, if the matrix is well conditioned, because $d$ is considered to be poly--logarithmic in the Hilbert--space dimension of the system.  This simple method works well in practice, but in cases where a precise estimate of the error is needed the numerical methods discussed in~\cite{ferrie_hpdregions_2013,granade_robust_2012} should be used.

A major drawback of this approach is that the likelihood function may, for certain experiments, be expensive to compute using classical computing resources.  Quantum Hamiltonian learning can resolve this problem if efficient quantum simulators exist for the class of models used in the inference procedure.

\section{Quantum Hamiltonian Learning}\label{sec:qhl}

\begin{figure}
\includegraphics[width=\linewidth]{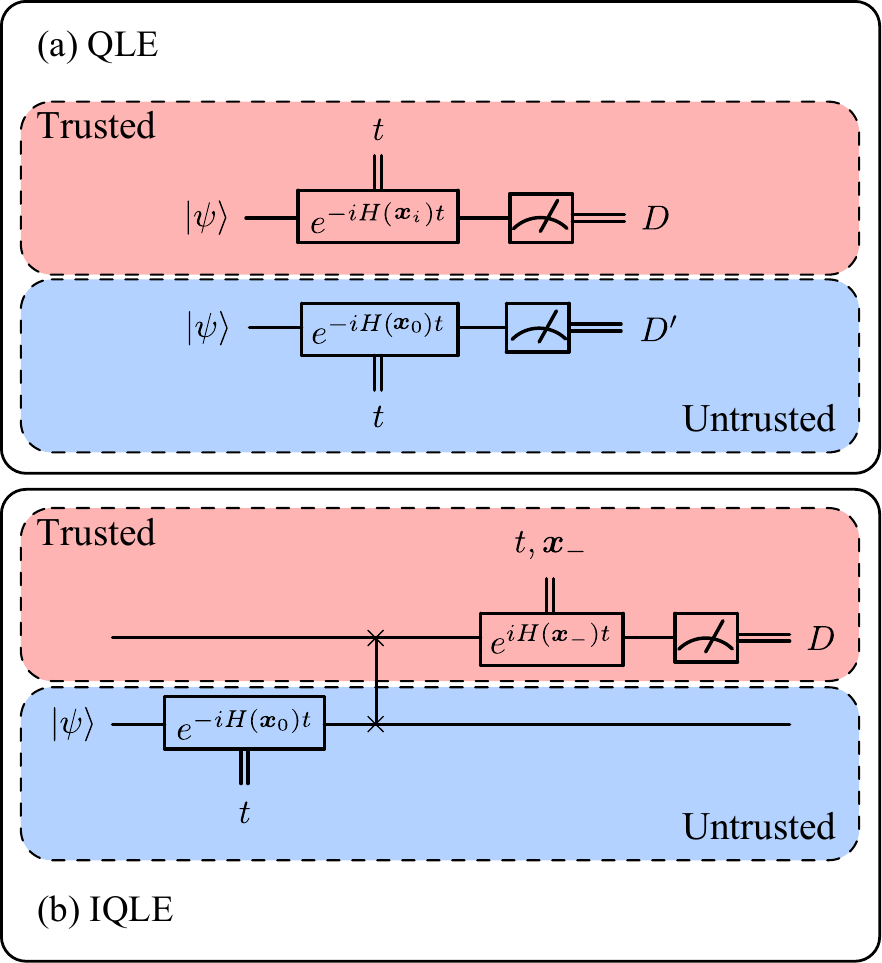}
  \caption{\label{fig:models} QLE and IQLE experiments: (a) \emph{quantum likelihood evaluation}, wherein the untrusted and trusted simulator act in parallel and the outputs are compared; and (b) \emph{interactive quantum likelihood evaluation}, wherein the state of the simulators is swapped and the trusted simulator attempts to invert the evolution.}
\end{figure}

The use of the likelihood function $\Pr(D|H(\vec{x}))\equiv\Pr(D|\vec{x})$ implies that the ability to infer $H$ is intimately connected to our ability to simulate dynamics
according to $H$, a fact that is clearly illustrated in Bayesian methods by~\eq{bayes}. This allows our implementation to be very general, and means that our estimation is always motivated by our knowledge of the underlying physics
of a system. 

Using classical simulation to implement evaluations of the likelihood function $\Pr(D|\vec{x})$ is typically extremely difficult for quantum systems.
For example, performing a dynamical simulation of random $12$-qubit quantum systems by direct exponentiation on a dual-processor Xeon E5-2630 workstation requires on average roughly
1300 seconds.
Given that the SMC approximation may require over $2000$ particles to achieve a good approximation, and that two hundred experiments may be needed to learn the Hamiltonian
parameters with reasonable accuracy, this would require approximately $16.5$ years of computational time on the workstation, despite these optimistic assumptions.
This is difficult, but not outside the realm of possibility for existing supercomputers.
If we wished to scale this up to $100$ qubits, as is applicable to current proposals for experimental quantum information processing devices \cite{britton_simulation_2012,richerme_trapped-ion_2013},
then roughly $5\times 10^{80}$ years would be required just to process the $20$ kilobits of data produced by the experiment.
Clearly, a better approach for characterizing the dynamics of mesoscopic quantum systems is needed.

A natural solution to this limitation is to use a quantum simulator to estimate the required likelihoods.  Efficient quantum simulators exist, in principle, for a wide range of physical systems~\cite{lanyon07,alan_qchem_2005,lanyon2010towards,gerritsma_diracsim_2010,kim_simulation_2010,britton_simulation_2012,richerme_trapped-ion_2013,wiebe2011simulating,BC12}.  Such simulators allow the user to draw a sample from $\Pr(D|\vec{x})$ using energy and time that scale at most polynomially with the number of interacting particles.  In contrast, the best known classical methods require time that scales polynomially in the Hilbert space dimension and hence cannot practically simulate generic large quantum systems (particular properties of certain systems can nonetheless be estimated using alternative techniques such as MPS, DMRG calculations or Lieb--Robinson bounds~\cite{daSilva_practical_2011}).  Such quantum simulators can take many forms.  It could be a universal quantum computer or it could be a special purpose analog quantum simulator.  Ultimately, the only things that matters are that it can, approximately, sample from $\Pr(D|\vec{x})$ for any $\vec{x}$ and that its state can be swapped with that in the uncharacterized quantum system.

Two ways have been proposed to use quantum information to accelerate Bayesian inference: QLE experiments and IQLE experiments.  A QLE experiment involves repeating the experiment that was performed on the uncharacterized quantum system a large number of times using a simulator for $H(\vec{x})$.  If datum $D$ was recorded for the experiment then $P(D|\vec{x})$ is set to be the fraction of times outcome $D$ is observed in the simulations.  IQLE experiments are similar except they involve swapping the quantum state out of the uncharacterized system using a noisy \swapgt~gate and then approximately inverting the evolution of the uncharacterized system by applying $e^{iH_- t}$. These methods are illustrated in \fig{models}.

\begin{algorithm*}[t!]
\caption{\label{alg:qhl} Quantum Hamiltonian learning algorithm.}
\rule{\linewidth}{1pt}
\begin{algorithmic}
\Require Particle weights $w_i$, $i \in \{1, \dots, M\}$, Particle locations $\vec{x}_i$,  $i \in \{1, \dots, M\}$, number of samples used to estimate probabilities $N_{\rm samp}$, total number of experiments used $N_{\rm exp}$, state preparation protocol for $\ket{\psi_0}$, protocol for implementing POVM $P$ such that $\ketbra{\psi_0}{\psi_0}$ is an element, resampling algorithm $R$.
\Ensure  Hamiltonian parameters $\vec{x}$ such that $H(\vec{x})\approx H(\vec{x}_{\rm true})$.

\vskip0.2em
\hrule
\vskip0.2em

\Function{QHL}{$\{w_i(D)\}$, $\{\vec{x}_i\}$, $N_{\rm samp}$, $N_{\rm exp}$, $P$, $R$}
  \For{$i \in 1 \to N_{\exp}$}
	\State Draw $x_-$ and $x'$ from $\Pr(\vec x)\colonequals w_i/\sum_i w_i $.\inlinecomment{Choose $x_-$ according to PGH}
	\State $t\gets 1/\|H(x)-H(x')\|$.\inlinecomment{Choose $t$ according to PGH}
	\State $D\gets $ measurement of $e^{iH(\vec{x_-})t}e^{-iH(\vec{x}_{\rm true})t}$ using $P$.\inlinecomment{Perform IQLE experiment on untrusted system.}
	\For{$j \in 1 \to M$}\inlinecomment{Compute likelihoods using trusted simulator}
	\State $p_j\gets 0$.
	\For{$k \in 1 \to N_{\rm samp}$}
		\State $D' \gets$ measurement of $e^{i H(\vec{x_-})t}e^{-i H(\vec{x}_j)t}\ket{\psi_0}$ using $P$.
		\If{$D' =D$}
			\State $p_j\gets p_j+1/N_{\rm samp}$.
		\EndIf
	\EndFor
	\EndFor
  \State $Z\gets \sum_{m=1,M}w_m p_m$.
  \State $w_i \gets w_i p_i/Z$.\inlinecomment{Perform  update.}
  \If{$1/(\sum_m w_i^{2})< M/2$}
	\State $(\{w_i\},\{\vec{x}_i\})\gets R(\{w_i\},\{\vec{x}_i\})$.\inlinecomment{Resample if weights are too small}
\EndIf
  \EndFor
  \State \Return $\sum_m p_m \vec{x}_m$\inlinecomment{Return Bayes estimate of $\vec{x}_{\rm true}$.}
\EndFunction
\end{algorithmic}
\rule{\linewidth}{1pt}
\end{algorithm*}

IQLE experiments have many advantages over QLE experiments~\cite{QHL}.  Firstly, if  {$H_-\approx H(\vec{x}_{\rm true})$, where $\vec{x}_{\rm true}$ are the true parameters,} and the noise is negligible then the simulation will approximately map $\ket{\psi}\mapsto\ket{\psi}$.  This is useful because it gives a firm condition to check to see if the current hypothesis about the Hamiltonian is correct.   In many cases, these benefits can outweigh the increased complexity of IQLE experiments and in particular, we will show that the Hamiltonian parameters can be learned even given realistic noise in the gate that swaps the states of the trusted simulator and the untrusted system.

IQLE experiments inherit their robustness in part from the use of the \emph{particle guess heuristic} (PGH), which is an adaptive method for choosing a reasonable experiment based on the current knowledge about the unknown quantum system.  The heuristic works by drawing two different ``particles'' from the prior distribution $\Pr(\vec{x})$, $\vec{x}_-$ and $\vec{x'}$.  The experimental time is chosen to be, $t=\|H(\vec{x}_-)-H(\vec{x}')\|^{-1}$.  This heuristic has several remarkable properties~\cite{QHL}:
\begin{enumerate}
\item The typical value of $t$ used for an experiment is the inverse of the \emph{current} uncertainty in the Hamiltonian.  Intuitively, this means that the guess heuristic will (on average) choose an evolution time that causes the majority of the potential models under consideration to have different dynamics.
\item If $\Pr(\vec{x})$ has converged to a unimodal distribution centered near $\vec{x}_{\rm true}$ then with high probability the measurement outcome will be $\ket{\psi}$, i.e. $|\bra{\psi}e^{iH(\vec{x_-})t}e^{-iH(\vec{x}_{\rm true})t}\ket{\psi}|^2\in O(1)$.
\item If $\|H(\vec{x}_{\rm true})- H(\vec{x}')\|$ is relatively large compared to $t^{-1}$ then the Loschmidt echo guarantees that $|\bra{\psi}e^{iH(\vec{x_-})t}e^{-iH(\vec{x}_{\rm true})t}\ket{\psi}|^2\in O\left(\frac{1}{{2^n}}\right)$ for almost all Hamiltonians (chosen, for example, uniformly over the Gaussian unitary ensemble of random Hamiltonians~\cite{ududec_equilibration_2013}).
\end{enumerate}
The key message from these results is that the PGH exploits the unitary (or approximately unitary) nature of time evolution to provide experiments that are likely to be informative.  In particular, the Loschmidt echo is exploited by IQLE experiments through the PGH to provide a test to determine with high probability whether the inferred Hamiltonian is close to the ``true'' Hamiltonian.  If the learning problem is well posed, then this allows the inference algorithm to learn that a constant fraction of model Hamiltonians, are closer to the true Hamiltonian than the other models.  This leads to a constant number of bits of information to be learned on average about the Hamiltonian per experiment, which leads to the uncertainty in the inference scaling like
\begin{equation}
\delta \sim A e^{-\gamma N},\label{eq:deltascale}
\end{equation}
where $N$ is the number of experiments performed and $\gamma$ is some constant which is independent of $N$.

\begin{figure*}[ht!]
\centering
\includegraphics[width=0.5\textwidth]{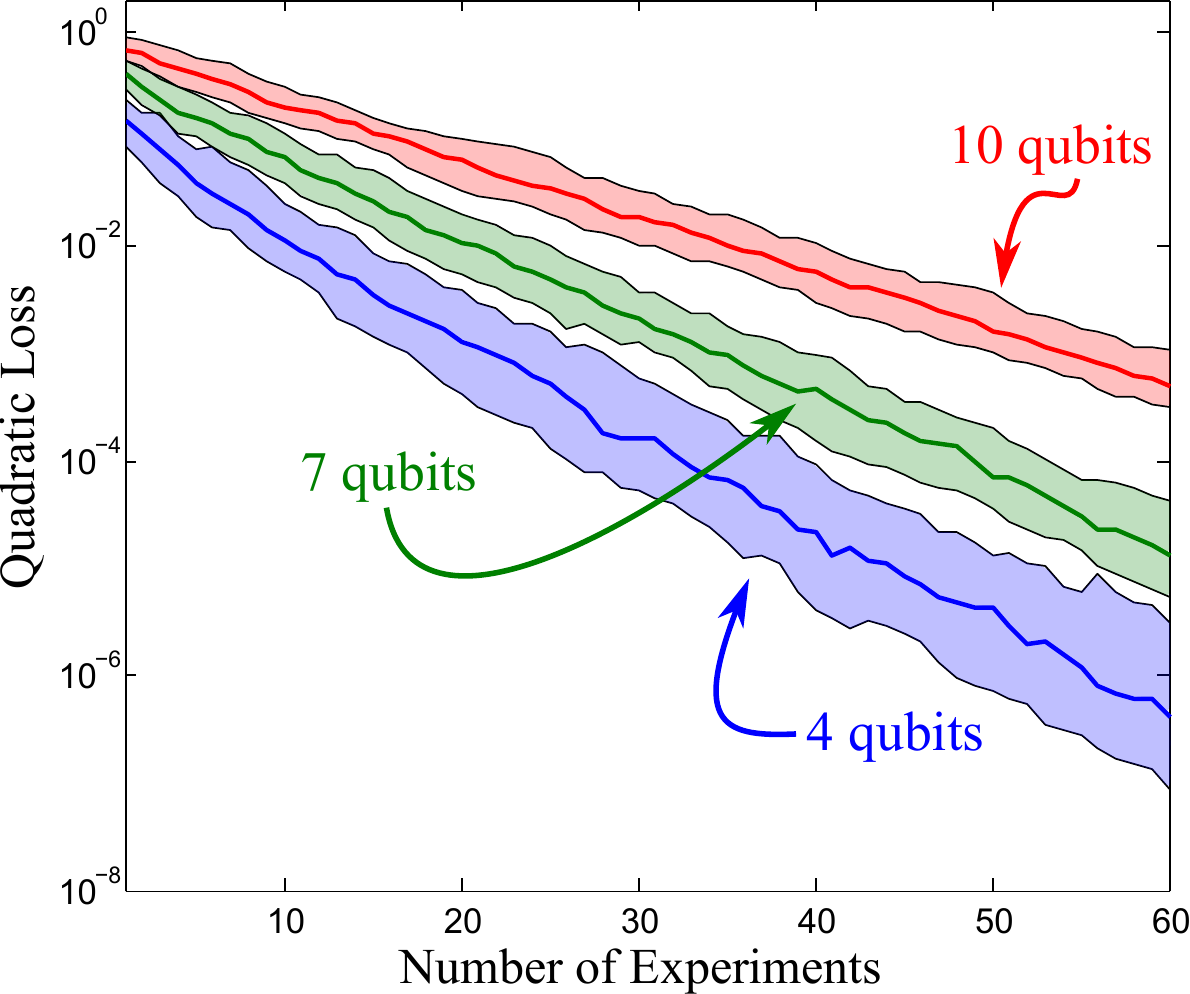}
\caption{Quadratic loss as a function of number of experiments for learning unknown Ising couplings.  The shaded regions represent a $50\%$ confidence interval for the quadratic loss and the solid line gives the median loss.}\label{fig:line_nonoise_plots}
\end{figure*}

The scaling in~\eq{deltascale} implies that the total computational time used scales as $t_{\rm total}\propto e^{\gamma N}$, which at first glance seems to suggest inefficiency but since the uncertainty also drops exponentially $N\propto \log(1/\delta)$ and hence $t\propto \delta^{-\gamma}$.  This scaling is comparable to that expected for Heisenberg--limited metrology in phase estimation protocols if $\gamma \approx 1$.  
In contrast, if the learning problem is less well posed and $\delta \sim N^{-\Gamma}$, it means that the total time required scales as $1/\epsilon^{\frac{\Gamma+1}{\Gamma}}$.  This yields similar scaling to shot noise limited metrology when $\Gamma=1$.

Putting everything together, we obtain an algorithm for performing an IQLE experiments, detailed in \alg{qhl}.  For more details on the resampling step and the SMC approximation see~\cite{granade_robust_2012}.  The method for performing QLE experiments is identical, except $H_-=0$ in those cases.

How do these methods compare to conventional approaches?  Techniques such as MPS tomography can be efficient for states with low bond--dimension and are error robust~\cite{Cramer2010Efficient}.  However, they are inefficient for systems that have high bond dimension, they use  potentially expensive (although efficient) tomographic processes  and can have exponentially worse scaling with the error than our approach.  Other approaches that such as the direct characterization method of~\cite{mohseni2006direct} apply to cases where the form of the true model is not known, require very little classical post processing and very few measurements if ensemble measurements are available.  On the other hand, these methods are inneficient; give exponentially worse scaling with the desired error if ensemble measurements are not available and do not exploit any prior information about the system.
Our previous work~\cite{granade_robust_2012} is inefficient, but scales exponentially better with the error tolerance than the aforementioned methods, can exploit prior information and is suitable in cases where single shot readout is used.  QHL therefore can combine the desirable properties of all of these methods at the price of using more sophisticated quantum resources.


\section{QHL Experiments Without Noise}\label{sec:nonoise}

Here, we examine QHL in the absence of noise to provide a basis of comparison for the performance of the algorithm when physically realistic noise is considered.  We will ignore the effects of sampling noise as it has already been studied in~\cite{QHL}, wherein the algorithm is shown to be robust against sampling errors given that a resampling algorithm is used and that experiments are chosen such that the majority of the outcomes do not have exponentially small likelihoods for typical particles in the SMC approximation to the prior distribution.

We consider a Hamiltonian of the form of~\eq{isingline} with unknown coefficients chosen to be uniformly distributed in the range $[-1/\pi,1/\pi]$.  We then apply the learning algorithm to a case where $20~000$ particles are used to describe the prior distribution and examine the error, given by the quadratic loss:
\begin{equation}
L(\vec{x}, \vec{x}_{\rm true}) = \sum_{j=1}^d (\vec{x}_j - {\vec{x}_{\rm true}}_{,j})^2,
\end{equation}
where $d=n-1$ is the number of model parameters.
We choose the Liu and West resampling algorithm with $a=0.9$ (see~\cite{granade_robust_2012} for more details) and, unless otherwise specified, the initial state $\ket{\psi}$ is chosen to be $\ket{+^{\otimes n}}$.  We examined the performance of the algorithm for $320$ different randomly chosen Ising models.  Although the model is not frustrated, previous work suggests that the absence of frustration does not qualitatively change the learning problem~\cite{QHL}.

We see from~\fig{line_nonoise_plots} that the quadratic loss shrinks exponentially with the number of experiments.  This is in agreement with prior results from~\cite{QHL}.  The learning rate, $\gamma$, which is found by fitting individual samples to $Ae^{-\gamma}$ agrees with an $O(1/d)$ scaling as shown in~\cite{QHL}.  Since $d=n-1$ for this model, such Hamiltonians should be easy to learn for IQLE experiments even in the limit of large $n$.  

\subsection{Non--Commuting Models}
{QHL is not limited to models with commuting Hamiltonians.  However, in the general non--commuting case,} it may be much more difficult to find appropriate initial states that maximize the information yielded by each experiment.  We illustrate this by applying QHL to a transverse Ising model of the form
\begin{equation}
H(\vec{x}) = \sum_{k=1}^n \vec{x}_k\sigma_x^{(k)} + \sum_{k=1}^{n-1} \vec{x}_{k+n} \sigma_z^{(k)} \otimes \sigma_z^{(k+1)}.
\end{equation}
The dynamics of the transverse Ising model are clearly much more rich than that of the Ising model and a na\"ive guess for an appropriate initial state/measurement operator for an IQLE experiment is unlikely to yield as much information as the choice of $\ket{\psi_0}=\ket{+}^{\otimes n}$ that was made in the prior example because that choice was motivated by the dynamics of the Ising interaction.

For example, a natural approach to solve such problems in NMR would be to use refocusing to suppress the $\sigma_z \sigma_z$ couplings while leaving the transverse field terms proportional to $\sigma_x$ by periodically applying $\pi)_x$ pulses to the system
being studied.
Such pulses can be designed in a broadband manner such that only rough knowledge of $H(\vec{x})$ is required to implement $\pi)_x$ pulses (see~\sec{model-selection} for more details on neglecting terms in the Hamiltonian).  After learning the these terms accurately, the interaction terms can be learned to much greater accuracy without suffering loss of contrast.
Nonetheless, we will show that in principle the QHL algorithm can be used directly to learn these couplings using maximally na\"ive experiments; specifically, we generate our initial states randomly in each experiment by applying a random series of local Clifford operations to each qubit, similar to~\cite{emerson_pseudo-random_2003,dankert2009exact}.

\begin{figure*}[t!]
\centering
\begin{minipage}[t]{0.49\linewidth}
\includegraphics[width=\linewidth]{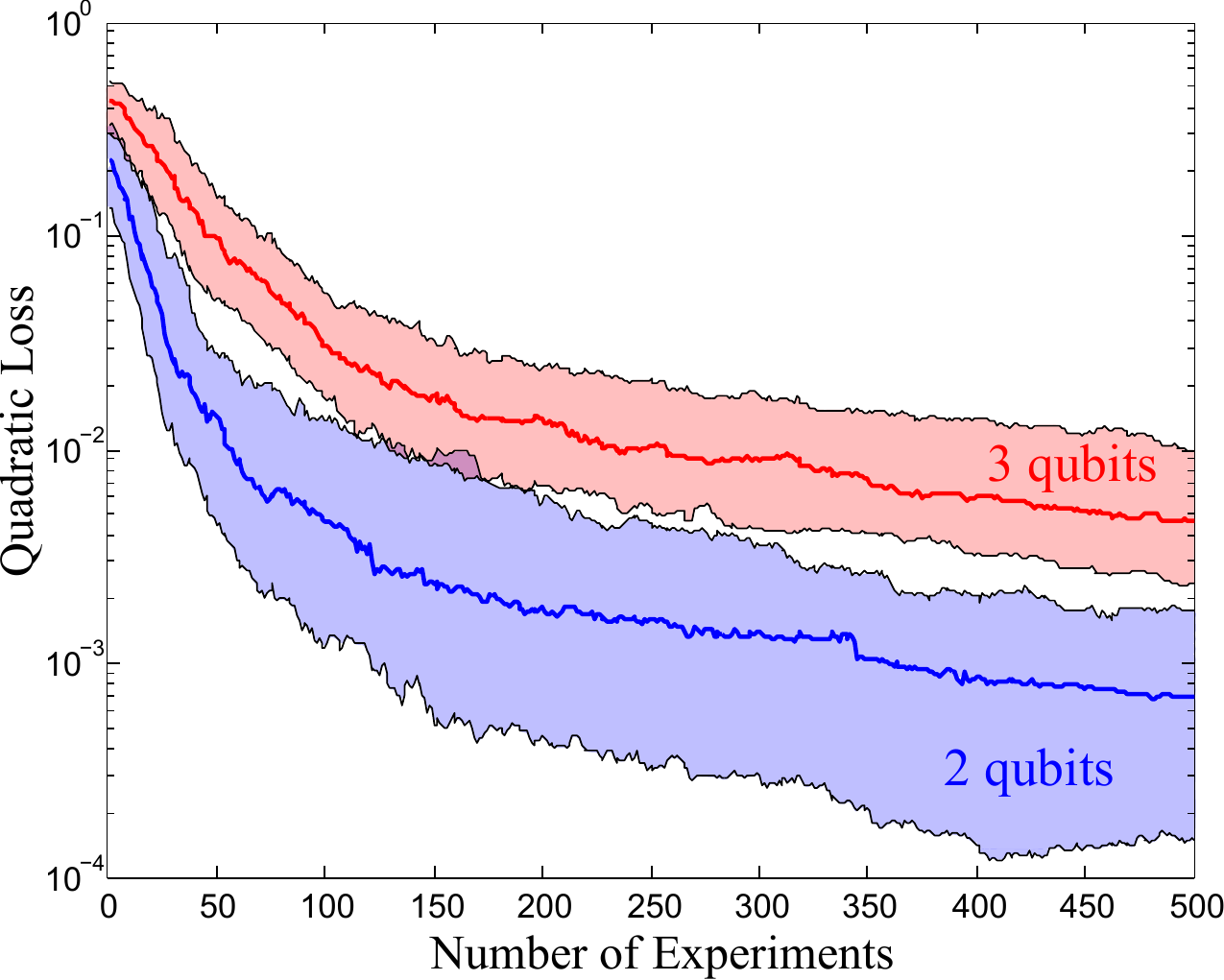}
\caption{Median quadratic loss as a function of the number of experiments for two-- and three--qubit transverse Ising Hamiltonians with chosen with $\vec{x}_k$ uniform in $[0,1]$.  The shaded regions give a $50\%$ confidence interval for the data.}\label{fig:noncommute}
\end{minipage}
\begin{minipage}[t]{0.49\linewidth}
\includegraphics[width=\linewidth]{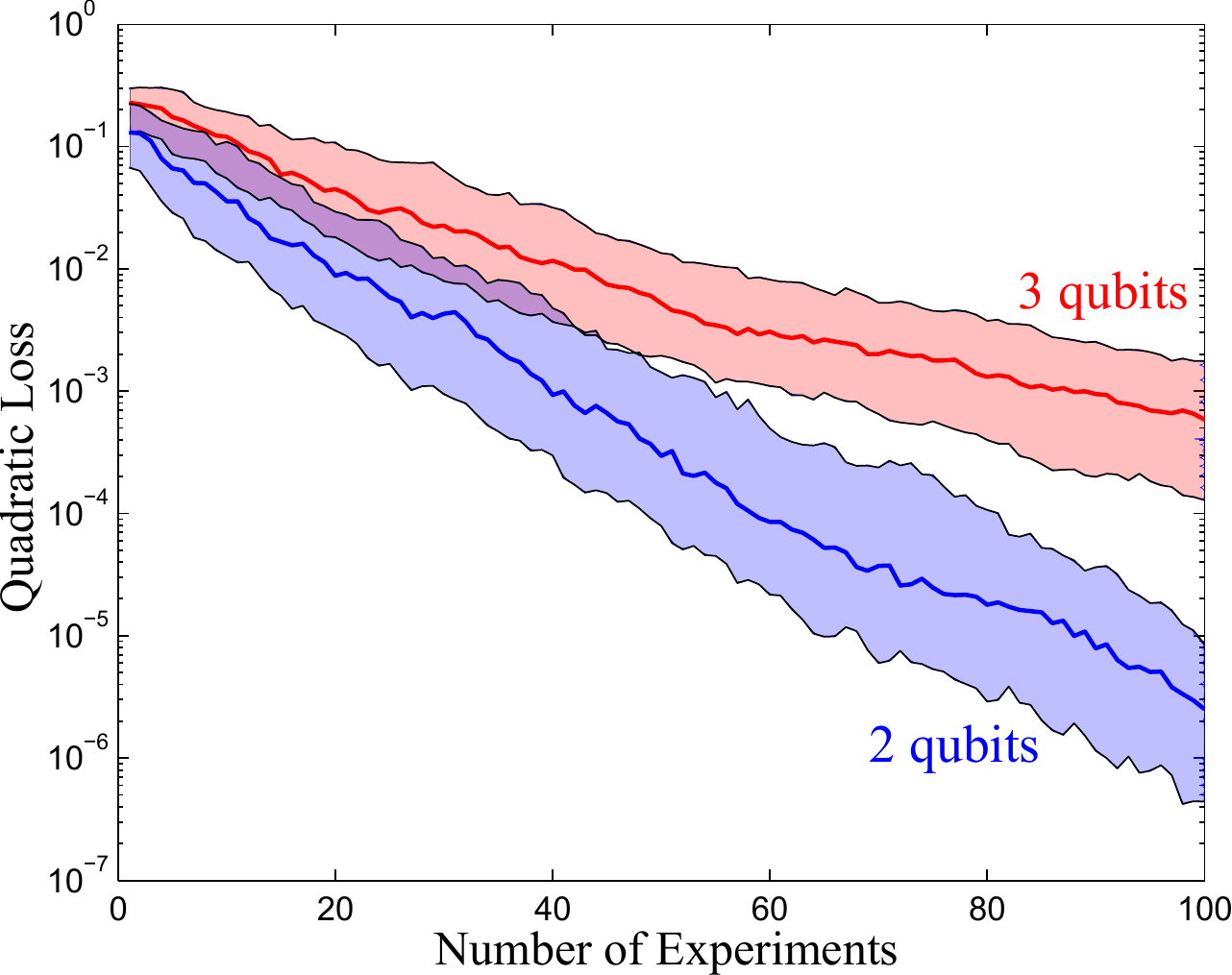}
\caption{Median quadratic loss as a function of the number of experiments for two and three--qubit translationally invariant transverse Ising model chosen with $\vec{x}_k$ uniform in $[0,1]$.  The shaded regions give a $50\%$ confidence interval for the data.}\label{fig:noncommute2d}
\end{minipage}
\end{figure*}

\fig{noncommute} shows that the QHL algorithm can continue to learn Hamiltonian parameters despite the fact that the model is non--commuting.  The data was collected using $5000$ particles, $160$ samples, and using the Liu--West resampling algorithm with $a=0.98$ for all numerical results in this section.  It is also worth noting that we are restricted to $2$ or $3$ qubit systems because (approximately) exponentiating non--commuting Hamiltonians is much more expensive classically than it is for commuting Hamiltonians.  The learning rate for short experiments is very rapid, whereas for later times the learning rate substantially slows.  This is expected because for short times $\exp(-i Ht)\approx \openone -i(\vec{x_1} \sigma_x^{(1)} + \vec{x_2} \sigma_x^{(2)}+ \vec{x_3} \sigma_z^{(1)}\otimes\sigma_z^{(2)})t$ and hence the single qubit and multi--qubit terms have a clear and separable effect on the experimental outcomes.  This results in rapid learning of these Hamiltonian parameters.  At later times, progress is substantially slower because the way that the Hamiltonian parameters affect the probabilities of different outcomes becomes less distinct.  Specifically, the scaling of the quadratic loss reduces to $\delta \propto 1/N$.  This results in a total simulation time that scales as $1/\epsilon^2$ which is comparable to shot--noise limited metrology.  We therefore see that in such cases the PGH alone is insuffcient to find highly informative experiments and intelligent choices of experiments and perhaps even local optimization (such as gradient descent optimization as per~\cite{granade_robust_2012}) become important for optimizing the performance of QHL.  

In essence, the slow learning rate of IQLE experiments in~\fig{noncommute} is a consequence of an approximate degeneracy that arises between the onsite terms and the interaction terms wherein the precise effect of a single onsite term becomes hard to resolve.  This raises the question of whether exponential scaling of the estimation accuracy can be restored if we break this degeneracy.  \fig{noncommute2d} shows that the answer to this question is ``yes''.  We break the degeneracy there by assuming that  the  interaction and onsite terms in the Hamiltonian are translationally invariant:
\begin{equation}
H(\vec{x}) = \vec{x_1}\sum_{k=1}^n \sigma_x^{(k)} + \vec{x_{2}}\sum_{k=1}^{n-1}  \sigma_z^{(k)} \otimes \sigma_z^{(k+1)}.
\end{equation}
In the case where $n=2$, this translationally invariant version has only one fewer parameter than the original Hamiltonian, and yet the performance differences in the learning algorithm are striking.  This emphasizes that finding good experiments that provide high contrast on the model parameters that we are trying to learn is crucial and that QHL in principle faces no difficulties in learning Hamiltonians with non--commuting terms.  We expect these qualitative features to remain the same even when $n>3$ because no new symmetries are introduced or broken as we scale up the system.

\section{Robustness of Quantum Hamiltonian Learning to Noise}\label{sec:noise}
In practice, swapping the quantum state out of the uncharacterized quantum system and into the trusted simulator will often be the most error prone step of an IQLE experiment.    The noise may be relatively small in some cases.  For example, in superconducting systems or scalable ion traps the trusted simulator could be part of the chip and the untrusted system could be another region that has not been adequately characterized.  The noise introduced by transferring the quantum state can be minimal since such architectures naturally permit state swapping.  On the other hand, noise in the \swapgt~operation could also be catastrophic in cases where the trusted simulator is not naturally coupled to the system, such as cases where flying qubits in the form of superconducting resonators or photons must be used as auxiliary resources to couple two simulators.  The inevitable question is: ``what level of noise can the learning algorithm sustain before it fails?''

We address the question by examining the performance of QHL using IQLE experiments for systems where the noise is known and the trusted simulator is capable of simulating the noise~\cite{LV01,wang2013solovay}.  We examine the performance of QHL theoretically and numerically for depolarizing noise, as well as physically realistic models of noise for quantum dots and superconducting circuits.  We will see that substantial depolarizing noise can be tolerated by the QHL algorithm and realistic noise models for existing swap gates similarly do not substantially impede learning.  We do not provide examples for non--commuting models here because such  numerical experiments are computationally expensive. 

\subsection{Depolarizing Noise}
\begin{figure*}
\begin{minipage}{0.45\textwidth}
\centering
\includegraphics[width=\textwidth]{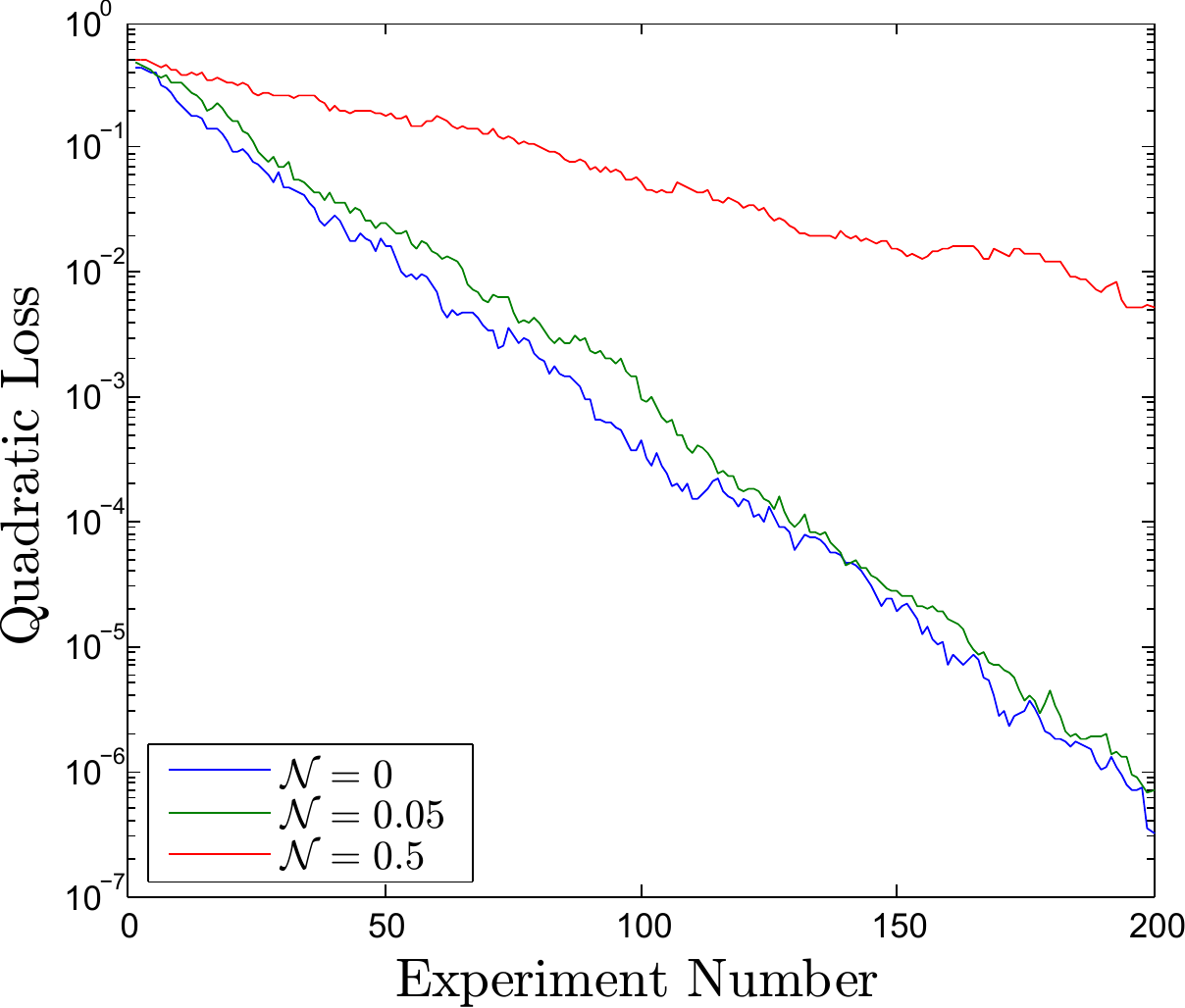}
\caption{The median quadratic loss plotted for IQLE experiments on the Ising Hamiltonian on a line with $4$ qubits for varying levels of depolarizing noise.}\label{fig:tpnoise}
\end{minipage}
\hspace{0.5 cm}
\begin{minipage}{0.45\textwidth}
\centering
\includegraphics[width=\textwidth]{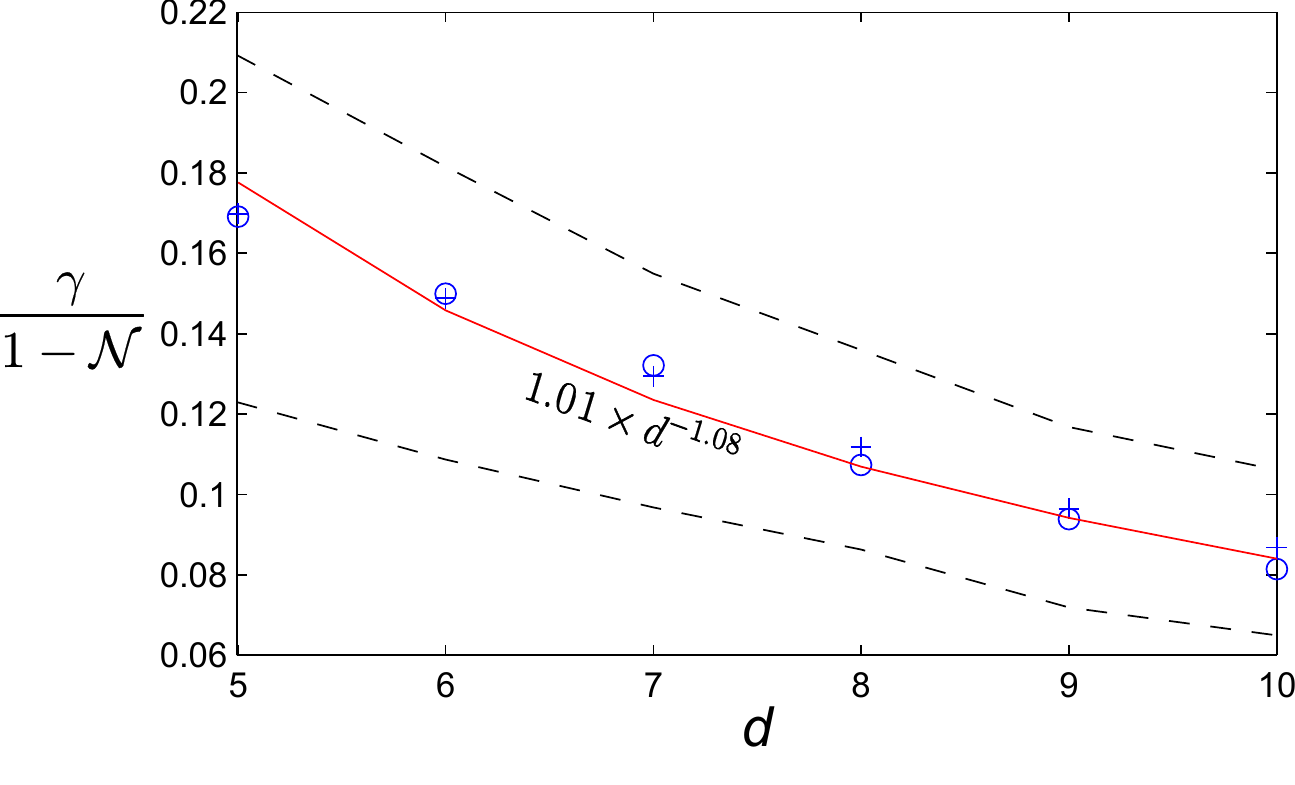}
\caption{The median value of $\gamma$ found by fitting  the quadratic loss of $200$ random local Hamiltonians to $Ae^{-\gamma N}$.  The dashed lines represent a $50\%$ condifence interval for the data, and the crosses and circles correspond to $\mathcal{N}=0.75/d$ and $\mathcal{N}=0.5/d$ respectively where $d$ is the number of model parameters.}\label{fig:linenoise}
\end{minipage}
\end{figure*}

The robustness of the learning algorithm to depolarizing noise arises similarly from the fact that a large number of particles are used in the SMC approximation, but also because of the fact that we assume that the strength of the depolarizing noise is known.  This robustness can be seen quite clearly in \fig{tpnoise}, where we show that $50\%$ depolarizing noise only slows the learning process down by a constant factor for random $4$-qubit Hamiltonians of the form of~\eq{isingline}. 
In contrast, $5\%$ depolarizing noise led to a negligible change in the scaling of the quadratic loss.

This is surprising at first glance because depolarizing noise implies that the final state of the system is \begin{equation}
\rho_{\rm final}=(1-\mathcal{N})\ket{\psi(t)}\!\bra{\psi(t)}+ \mathcal{N} \openone/2^n,\label{eq:noisedef}
\end{equation}
 for a known value of $\mathcal{N}$. However, this is not a substantial problem if $\mathcal{N}$ is known because we can calculate likelihoods for quantum experiments with a fixed amount of depolarizing noise.  This, in effect, means that the worst thing that well characterized depolarizing noise can do is reduce the visibility of an experiment, which in turn simply slows the learning process by a factor no worse than $\mathcal{N}^2$ \cite{ferrie_how_2012}.  In fact, the rate at which IQLE experiments learn the unknown parameters will typically only be slowed down by a rate proportional to $\frac{1}{1-\mathcal{N}}$ compared to the noise--free case.

It is worth noting that the depolarizing channel commutes with all the operations in an IQLE experiment.  This means that it does not matter when the depolarizing noise is introduced to the system, unlike the other noise models that we will consider.
For convienence, then, in our numerical simulations, we have applied the depolarizing noise at the end, represented as an effective visibility as given by \eq{noisedef}.

Let us consider an IQLE experiment where the measurement results are coarse grained into two outcomes, $\psi_0$ and its orthogonal compliment $\psi_0^\perp$.  Then Bayes' theorem states that the expected update to the prior distribution of a given experiment is
\begin{widetext}
\begin{equation}
\mathbb{E}_{y\in \{\psi_0, \psi_0^\perp\}}\left(\frac{\Pr(\vec{x}|y)}{\Pr(\vec{x})}\right)=\frac{\Pr^2(\psi_0|\vec{x})}{\sum_j \Pr(\psi_0|\vec{x}_j)\Pr(\vec{x}_j)}+\frac{\Pr^2(\psi_0^\perp|\vec{x})}{\sum_j \Pr(\psi_0^\perp|\vec{x}_j)\Pr(\vec{x}_j)}.\label{eq:expectupdate}
\end{equation}
If in the absence of noise, $\Pr(\psi_0|\vec{x})=A$ and $\Pr(\psi_0^\perp|\vec{x})=1-A$ then~\eqref{eq:noisedef} gives us that for any~\vec{x}
\begin{align}
\Pr(\psi_0|\vec{x})&= A(1-\mathcal{N})+\frac{\mathcal{N}}{2^n}\label{eq:goodprob}\\
\Pr(\psi_0^\perp|\vec{x})&= (1-A)(1-\mathcal{N})+\frac{\mathcal{N}(2^n-1)}{2^n}\label{eq:badprob}
\end{align}
 in the presence of depolarizing noise.

We then find the following by substituting~\eq{goodprob} and~\eq{badprob} into~\eq{expectupdate} and assuming that $\min_j(1-A_j)\gg \mathcal{N}/(1-\mathcal{N})$ along with $\mathcal{N} \ll 2^n$:
\begin{align}
\mathbb{E}_{D\in \{\psi_0, \psi_0^\perp\}}\left(\frac{\Pr(\vec{x}|D)}{\Pr(\vec{x})}\right)&=\frac{\left(A(1-\mathcal{N})+\mathcal{N}/2^n \right)^2}{\sum_j (A_j(1-\mathcal{N}) +\mathcal{N}/2^n)P(\vec{x}_j)}+\frac{\left((1-A)(1-\mathcal{N})+\frac{\mathcal{N}(2^n-1)}{2^n} \right)^2}{\sum_j ((1-A_j)(1-\mathcal{N}) +\mathcal{N}(2^n-1)/2^n)P(\vec{x}_j)}\nonumber\\
&\sim (1-\mathcal{N})\left(\frac{A^2}{\sum_j A_j \Pr(\vec{x}_j)}+\frac{(1-A)^2}{\sum_j (1-A_j)\Pr(\vec{x}_j)}\right).
\end{align}
\end{widetext}
Therefore, under these assumptions, the expected relative impact of an observation on the posterior probability $\Pr(\vec{x}|D)$ is a factor of $1-\mathcal{N}$ smaller than would be expected in the absence of noise.  This in turn suggests that the learning rate, as parameterized by $\gamma$ scales like $1-\mathcal{N}$ in the presence of depolarizing noise.  This shows theoretically that small amounts of depolarizing noise will not be sufficient to destabilize QHL.

\fig{linenoise} shows that the median value of $\gamma$ found by fitting the quadratic loss for $200$ random Ising models of the form of~\eq{isingline}, for varying $d$, to $Ae^{-\gamma N}$ scales as $1-\mathcal{N}$.  Twenty--thousand particles were used for these experiments.  The constant $\gamma$ clearly represents a characteristic timescale for the learning problem, and hence it is clear that the learning rate is slowed by a factor of $(1-\mathcal{N})$ for these problems.  This agrees with our prior theoretical expectations in the limit where $\Pr(D|\vec{x}_j)\gg \mathcal{N}/(1-\mathcal{N})$.  It is further worth noting that the learning rate scales roughly as $d^{-1}$, which suggests that the cost of Hamiltonian learning scales efficiently with the number of qubits in the chain, as also noted in~\cite{QHL}.

This shows that our method is robust to the presence of a well characterized source of depolarizing noise.  It is worth mentioning, however, that the reduced visibility imposed by the depolarizing noise may be especially problematic for QLE experiments since the distribution of outcomes tends to be much flatter in such cases than the corresponding outcome distributions for IQLE experiments.  This further underscores the utility of using the trusted simulator in an interactive fashion in such learning protocols.

\subsection{Realistic Models for $\swapgt$~Gate Errors}

While the above argument lets us reason analytically about the effects of depolarizing noise on inference, in practice
the implementation of a $\swapgt$ gate need not admit as simple a description as that. To remedy this, a more complete model of the errors in a $\swapgt$ implementation can be incorporated 
into our IQLE simulations.
In particular, starting from the cumulant expansion \cite{kubo_generalized_1962},
we can simulate the effects of stochastic processes in the
environment, open quantum dynamics and the limited fidelity of a particular shaped pulse sequence derived by optimal control
theory \cite{khaneja_optimal_2005}.
The cumulant expansion generalizes the Magnus expansion \cite{magnus_exponential_1954} to incorporate the effects of
stochastic operators, and has been used in quantum information to design control methods
that are robust to stochastic fields \cite{cappellaro_principles_2006}.
Numerically evaluating a truncation of the cumulant expansion then gives us a \emph{superoperator} that
describes the action of the $\swapgt$ gate, so that we can reproduce its effect on the trusted simulation alone by engineering
noise on that system \cite{fortunato_implementation_2002}.

Concretely, the cumulant expansion provides a solution to the ensemble-average time-ordered exponential
\begin{equation}
    \hhat{S}(t) = \left\langle \T\exp\left(\int_0^t \hhat{G}(t) \dd t\right) \right\rangle,
\end{equation}
where $\hhat{G}(t)$ is a stochastic and time-dependent operator in $\Lin(\Lin(\H))$ (commonly denoted as a superoperator),
such that
\begin{equation}
\hhat{G}(t)[\rho] = -i[H(t), \rho] + \hhat{D}[\rho]
\end{equation}
for a Hamiltonian operator
$H$ and a dissipative map $\hhat{D}$.
That is, $\hhat{G}(t)$ is a superoperator implementing the adjoint map $\ad H(t)$
together with the generator of a quantum dynamical semigroup.

Given that $\hhat{G}$ is a linear operator, we can represent it as a matrix acting on
$\Lin(\H)$, the elements of which are \emph{vectors} representing \emph{operators}
in the original vector space $\H$. A convienent choice for such vectorizations
$\Sket{\rho}$ is to stack the columns of $\rho$ to make a vector--- for example,
\begin{equation}
    \Sket{\left(\begin{matrix}a & b \\ c & d\end{matrix}\right)} = \left(\begin{matrix}a \\ b \\ c \\ d\end{matrix}\right).
\end{equation}
More generally, $\Sket{\ket{i}\bra{j}} = \bra{j}^{\T} \otimes \ket{i}$ in this convention.

Using this formalism, $\hhat{S}(t)$ is seen to be a propagator acting on $\Lin(\H)$ that represents the
effect of the stochastic process described by $\hhat{G}(t)$ on vectorizations of mixed states $\Sket{\rho}$.
Truncating the cumulant expansion at the second order,
\begin{align}
    \hhat{S}(t) & = \exp(\hhat{K}_1 + \hhat{K}_2), \text{where} \\
    \nonumber \hhat{K}_1  & = \frac{1}{t} \int_0^{t} \dd t_1 \left\langle \hhat{G}(t_1) \right\rangle \\
    \nonumber \hhat{K}_2  & = \frac{1}{t^2} \T \int_0^{t} \dd t_1 \int_{0}^{t} \dd t_2 \left\langle\hhat{G}(t_1)\hhat{G}(t_2)\right\rangle - \hhat{K}_1^2.
\end{align}
These integrals can readily be numerically computed by characterizing the stochastic process
$\hhat{G}$ in terms of a correlation function, as is discussed at greater length in
\cite{puzzuoli_tractable_2013}.

Applying this expansion to the problem of simulating realistic errors in coupling the trusted and untrusted simulators,
we start with the models of a superconducting and quantum dot systems described by
Puzzuoli et al \cite{puzzuoli_tractable_2013}, using the parameters
described in \app{realistic-sim-params}. Next, we use a gradient ascent optimization method known as the GRAPE algorithm \cite{khaneja_optimal_2005} to design a $\swapgt$
implementation using the controls admitted by each of these systems.  We also consider two superconducting models whose noise strength has been substantially
increased, resulting in lower-fidelity implementations for comparison.
The quantum dots $\swapgt$ implementation uses an XY4 sequence \cite{gullion_extended_1989} to decouple from the environment. In the
superconducting model, we consider both an XY4 and a ``primitive'' (that is, a single pulse found using optimal control theory via GRAPE)
implementation for the lowest-noise case and the ``primitive'' implementation only for the other two models.

We then find the noise
map $\Lambda_{\text{noise}}$ for the cumulant-simulated superoperator $\hhat{S}_{\swapgt}$ for each swap gate used the IQLE experiment (see \fig{models}).
In particular, we note that the action of the \swapgt~gate on the input state $\rho = \ketbra{\psi}{\psi}$ is given by
\begin{equation}
    \rho \mapsto \Tr_{\text{untrusted}} \left( \hhat{S}_{\swapgt} [ \rho \otimes \ketbra{0}{0} ] \right),
\end{equation}
where $\ketbra{0}{0}$ is the initial state of the trusted quantum simulator. By representing the state preparation and partial trace
as non-rectangular superoperators, we have that in the supermatrix representation,
\begin{equation}
    \Lambda_{\text{noise}} = \hhat{S}_{\Tr_{\text{untrusted}}} \circ \hhat{S}_{\swapgt} \circ \hhat{S}_{\text{prep}}.
\end{equation}
Note that even though there is no noise in either the trace or the preparation,
it is convienent to keep with superoperators so that composition of maps
is represented by simple matrix multiplication.

For a single qubit, we can easily express these superoperators in the column-stacking basis
of $\Lin(\Lin(\H))$ as
\begin{align}
    \hhat{S}_{\text{prep}} & = \left(\begin{smallmatrix}
        1 & 0 & 0 & 0\\
        0 & 0 & 0 & 0\\
        0 & 1 & 0 & 0\\
        0 & 0 & 0 & 0\\
        0 & 0 & 0 & 0\\
        0 & 0 & 0 & 0\\
        0 & 0 & 0 & 0\\
        0 & 0 & 0 & 0\\
        0 & 0 & 1 & 0\\
        0 & 0 & 0 & 0\\
        0 & 0 & 0 & 1\\
        0 & 0 & 0 & 0\\
        0 & 0 & 0 & 0\\
        0 & 0 & 0 & 0\\
        0 & 0 & 0 & 0\\
        0 & 0 & 0 & 0
    \end{smallmatrix}\right) \\
\intertext{and}
    \hhat{S}_{\Tr_{\text{untrusted}}} & = \left(\begin{smallmatrix}
        1 & 0 & 0 & 0 & 0 & 0 & 0 & 0 & 0 & 0 & 1 & 0 & 0 & 0 & 0 & 0\\
        0 & 1 & 0 & 0 & 0 & 0 & 0 & 0 & 0 & 0 & 0 & 1 & 0 & 0 & 0 & 0\\
        0 & 0 & 0 & 0 & 1 & 0 & 0 & 0 & 0 & 0 & 0 & 0 & 0 & 0 & 1 & 0\\
        0 & 0 & 0 & 0 & 0 & 1 & 0 & 0 & 0 & 0 & 0 & 0 & 0 & 0 & 0 & 1
    \end{smallmatrix}\right).
\end{align}
The multiple-qubit superoperators are found from these single qubit operators using the techniques described in
\cite{wood_tensor_2011}.

While simulating or characterizing a $\swapgt$~gate in this manner is not in general tractable, recent work
demonstrates that we can obtain an \emph{honest approximation} to gates such as the $\swapgt$~gate
that are restricted to a subclass of efficiently simulatable channels, but which only exaggerate the error
\cite{puzzuoli_tractable_2013}.
In the case of SMC, this exaggerated error manifests as an additional source of sampling error, such that
we can make a tradeoff between the error within the SMC procedure and the accuracy with which we model
quantum couplings between the trusted and untrusted registers.  Nonetheless, the cost of these simulations
limits our numerics to only two qubits.

In addition to allowing for reduction in the resources required for this secondary characterization
task, honest approximation allows us to reduce the simulation resources needed to model an IQLE experiment
entirely in the trusted register, such that a full open-system simulation need not be necessary.
For instance, if we wish to represent the $\swapgt$~gate by a Pauli channel immediately following
the $\swapgt$, we can use the Puzzuoli algorithm to find the probabilities with which we should insert
perfect single-qubit gates into the trusted register so as to honestly approximate the dynamics of the actual
coupling.

The resultant density operator can be computed for an IQLE inference procedure via,
\begin{align}
  \begin{split}
    \rho(t)        & = e^{i H_- t} \Lambda_{\text{noise}}\biggr[ \\
    & \qquad e^{-iH(\vec{x})t}\ketbra{+^{\otimes 2}}{+^{\otimes 2}} e^{iH(\vec{x})t}\biggr]e^{-iH_{-}t},
  \end{split}\\
\intertext{or equivalently, using the vectorization $\Sket{\rho}$,}
  \begin{split}
    \Sket{\rho(t)} & = e^{i \hhat{G}_- t} \circ \hhat{S}_{\Tr_{\text{untrusted}}} \circ \hhat{S}_{\swapgt} \circ \hhat{S}_{\text{prep}} \\*
    & \qquad \circ e^{-i\hhat{G}(\vec{x})t} \Sket{\ketbra{+^{\otimes 2}}{+^{\otimes 2}}},
  \end{split}
\end{align}
where $\Lambda_{\text{noise}}$ is the result of a cumulant simulation, and
where the Hamiltonian used is the $J$-coupling between a pair of qubits,
\begin{equation}
    H = J \sigma_z^{1} \sigma_z^{2},
\end{equation}
equivalent to the $n=2$ line model above, and where the $\hhat{G}(\vec{x})\Sket{\rho} = \Sket{[H(\vec{x}), \rho]}$,
representing that no noise acts on the system during free evolution before and after
the imperfect $\swapgt$ gate. We let $\hhat{G}_- = \hhat{G}(\vec{x}_-)$ in analogy
to our notation for $H$.

\begin{figure*}[!]
\centering
\includegraphics[width=0.95\linewidth]{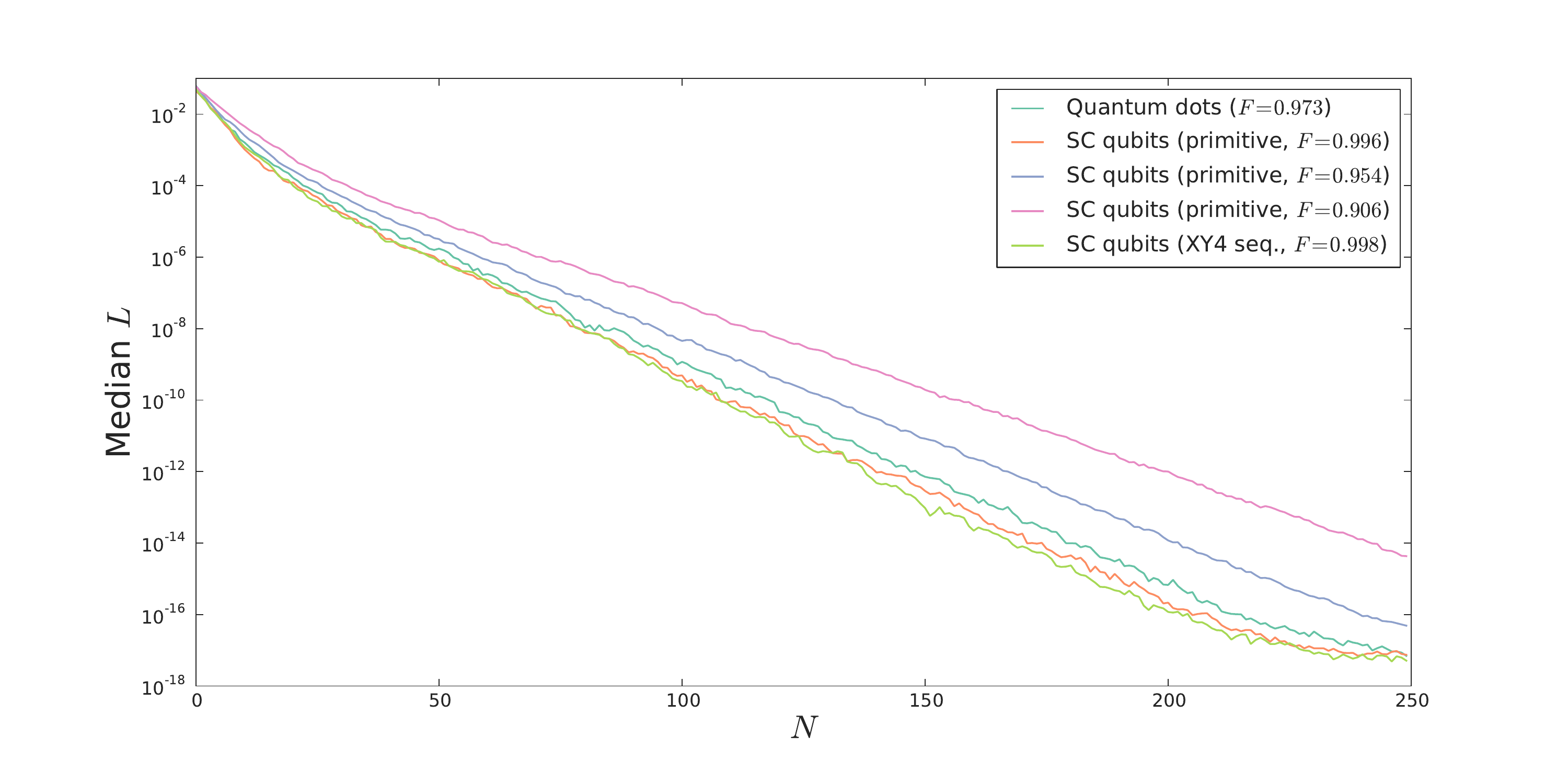}
\caption{The median value of the loss $L$ in estimating the $J$-coupling between two qubits as a function of the number of measurements $N$ performed,
for each of the five physical models of the $\swapgt$ gate considered.
The fidelity $F$ of each implemented $\swapgt$ is shown inset.}
\label{fig:realistic-median}
\end{figure*}

\fig{realistic-median} shows that IQLE experiments continue to gain information at an exponential rate for these realistic levels of noise.  The learning is so rapid, that after approximately $200$ experiments the unknown value of $J$ will be known to $6$ to $7$ digits of accuracy despite the fact that the \swapgt~gate infidelities are $4$ or $5$ orders of magnitude greater than these uncertainties.   This robustness largely arises because the model knows the noise model for the \swapgt~gate.  If it did not, then we would not see such large separations between the magnitudes of the uncertainties in $J$  and the gate infidelities.  We explore this point in more detail in the next section.  It is worth noting before proceeding, that in principle we do not need to know the precise noise model of the \swapgt~gate before performing the experiment: it can be learned simultaneously with $J$ using the exact same approach (see~\cite{granade_robust_2012} for such an example).

\section{Robustness of Algorithm to Errors in Model}
\label{sec:model-selection}
When modeling physical systems, it is usually not convenient to include every possible interaction that could exist in the system.  For example, in spin systems with dipolar coupling it is common to neglect interactions between distant particles in the system because such interactions decay rapidly with distance.  This raises a problem for quantum Hamiltonian learning: it is quite likely that the untrusted quantum system contains couplings that are not modeled by the trusted simulator.  It is therefore important to show that QHL will remain stable and continue to infer the best possible model in spite of the fact that the set of allowed models does not contain the true Hamiltonian.  We show here that small discrepancies between the model used in the trusted simulator and the true model for the untrusted system are not catastrophic for QHL; infact, QHL continues to learn $H$ until saturating at a  level of uncertainty that scales at most linearly with the number of neglected terms.

It is shown in~\cite{NC00} that for any two Hamiltonians $H$ and $\tilde{H}$
\begin{equation}
\|e^{-iHt} - e^{-i\tilde H t}\|\le \|H-\tilde{H}\|t.
\end{equation}
This implies that if the Hamiltonain $H$ is used to model the Hamiltonian $\tilde{H}$ then the error in the likelihood function obeys
\begin{align}
\begin{split}
\Delta \Pr(D)&:= |\bra{D} e^{-iHt}\ket{\psi_0}|^2-|\bra{D} e^{-i\tilde Ht}\ket{\psi_0}|^2\\
&\le |\bra{D} e^{-iHt}- e^{-i\tilde Ht}\ket{\psi_0}|^2\\
&\le \|H-\tilde{H}\|^2t^2.\label{eq:Hstab}
\end{split}
\end{align}
Equation~\eqref{eq:Hstab} implies that the error due to using an approximate Hamiltonian model is negligible provided $\|H-\tilde{H}\|\ll t^{-1}$.
Our use of the particle guess heuristic implies that the time chosen is (with high probability) approximately the reciprocal of the uncertainty of the uncertainty in the Hamiltonian (i.e. $t\propto \Delta H^{-1}$).  The use of an inexact model therefore is not problematic for the inference algorithm unless
\begin{equation}
\|H-\tilde{H}\|\approx \Delta H.
\end{equation}
In particular, if we parameterize the Hamiltonians via $\vec x\in \mathbb{R}^{d}$ as $H(\vec{x})$, then it is sufficient to assert that
\begin{equation}
\min_{\vec x} \|H(\vec x) - \tilde{H}(\vec x)\| \ll \Delta H.\label{eq:errstab}
\end{equation}
It is, however, often sufficient in practice to assert that the expectation value over all particles is sufficiently small compared to $\Delta H$.

Also note that if the terms $\mathcal{H}_1,\ldots,\mathcal{H}_R$ are neglected from the Hamiltonian model then $\min_{\vec x} \|H(\vec x) - \tilde{H}(\vec x)\| \le R \max_{j=1,\ldots,R} \|\mathcal{H}_j\|$, which implies that the use of an inexact model will not be problematic if
\begin{equation}
\frac{\Delta H}{R} \gg \max_{j=1,\ldots,R} \|\mathcal{H}_j\|.
\end{equation}
This implies that the point at which the algorithm ceases to learn varies at most linearly with $R$ (assuming $\|\mathcal{H}_j\|$ is independent of $R$).  Since $R$ will typically vary polynomially with the number of interacting particles in a system, our algorithm remains tractable for physically motivated high-dimensional systems.


\begin{figure*}[t!]
\begin{minipage}[t]{0.49\linewidth}
\centering
\includegraphics[width=\linewidth]{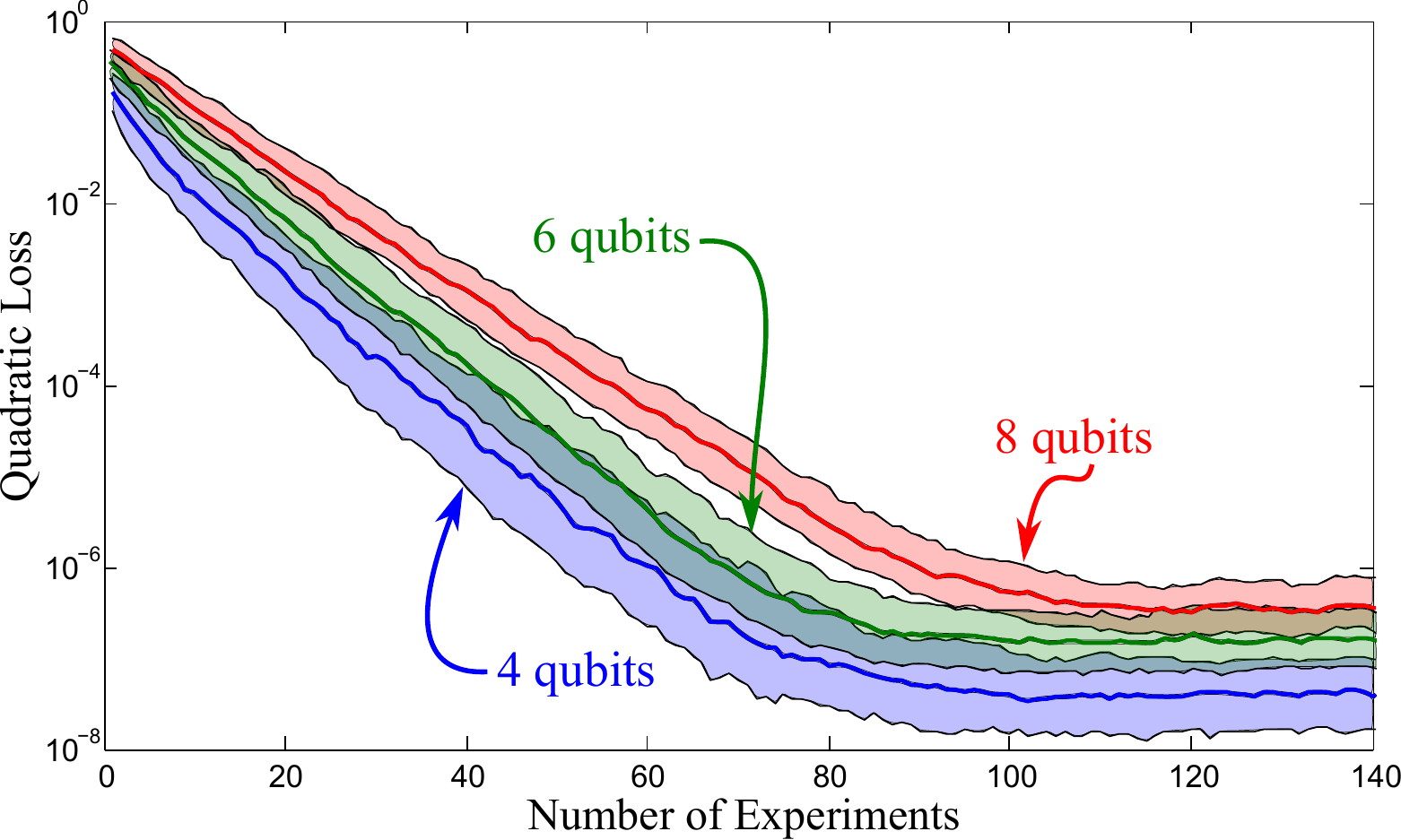}
\caption{The performance of QHL for the case where the trusted simulator uses an Ising model on the line given that the true Hamiltonian is an Ising model on the complete graph with non--nearest neighbor interactions on the order of $10^{-4}$ and nearest neighbor interactions on the order of $0.5$.}\label{fig:badmodel}
\end{minipage}
\begin{minipage}[t]{0.49\linewidth}
\centering
\includegraphics[width=\linewidth]{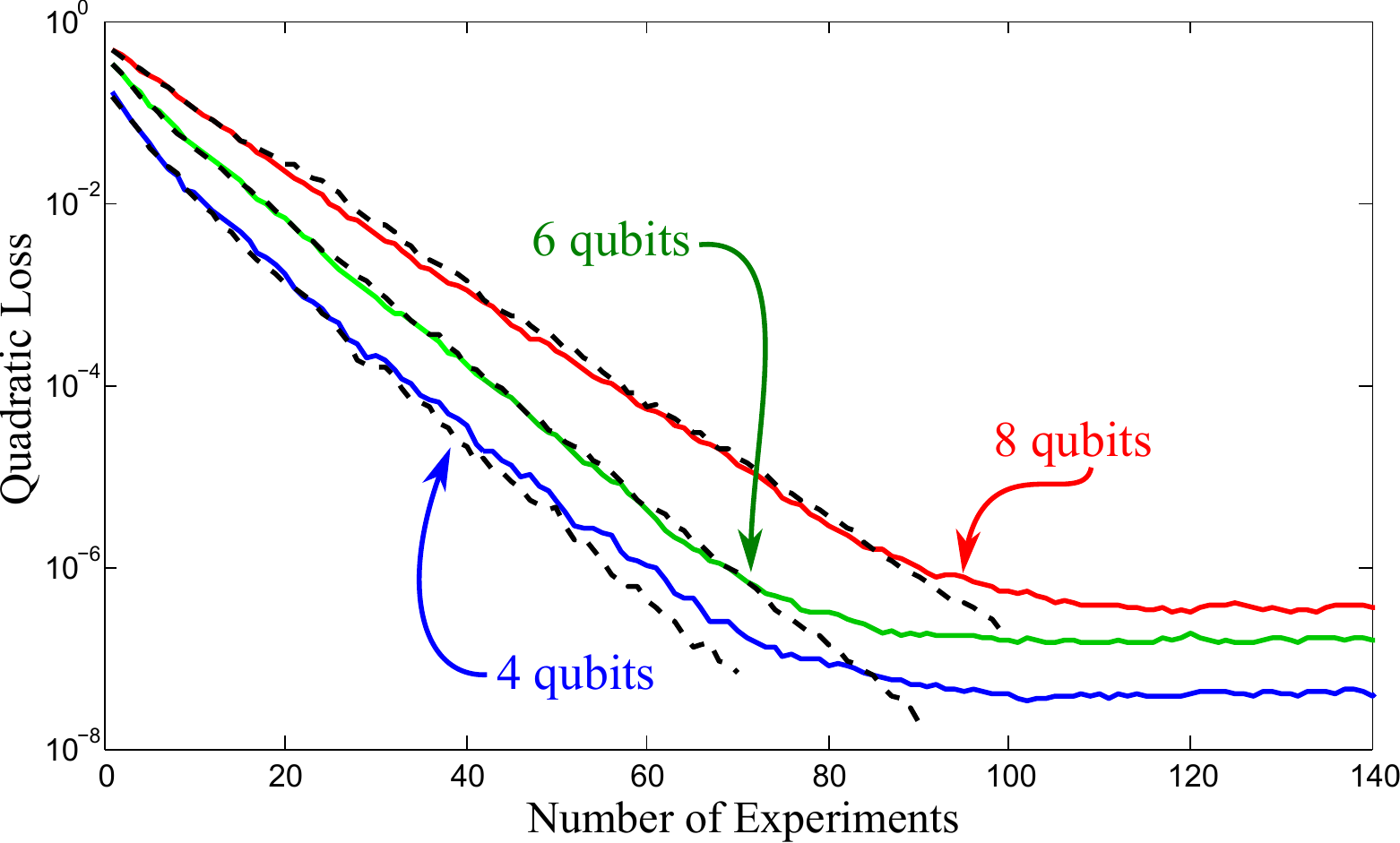}
\caption{As~\fig{badmodel} with the addition of data (dotted lines) for the performance when the true Hamiltonian only has nearest--neighbor couplings on a line.  In contrast, the true Hamiltonian in~\fig{badmodel} contains small non--nearest neighbor couplings.}\label{fig:noisecompare}
\end{minipage}
\end{figure*}

We see this behavior clearly illustrated in \fig{badmodel} where we examine the performance of QHL given that an inexact model is used for the unknown Hamiltonian.  In particular, we take

\begin{align}
\begin{split}
H &= \sum_{i=1}^{n-1} a_i \sigma^{z}_i \sigma^z_{i+1}\\ 	
\tilde{H} & = \sum_{i=1}^{n-1} a_i \sigma^{z}_i \sigma^z_{i+1} + \sum_{i=1}^{n-1}\sum_{j=i+2}^{n} b_{i,j} \sigma^{z}_i \sigma^z_{j}\label{eq:HH'}
\end{split}
\end{align}
The coupling constants $a_i$ are each chosen uniformly from the interval $[-1/2,1/2]$ and the coupling constants $b_{i,j}$ are each chosen according to a Gaussian distribution with mean zero and standard deviation $10^{-4}$.  In this case the models have different dimension, so we compute the quadratic loss by square error in the inferred values of the $a_i$ only.  

Note that in practical cases, such as those based on dipolar Hamiltonians, next--nearest neighbor interactions are often on the order of $10^{-1}$.  We use $10^{-4}$ to illustrate the qualitative difference between the regime in which the algorithm is learning the Hamiltonian and the region where learning ceases by creating a stronger separation between included and neglected terms.

We note that for the data considered in~\fig{badmodel} that learning proceeds at an exponential rate until saturating at a value that is approximately on the order of $\|H-\tilde{H}\|^2$.  This shows that QHL is still valuable in cases where an inexact model is used by the simulator, which further underscores the utility of this procedure in finding Hamiltonian models for unknown quantum systems.  \fig{noisecompare} shows us that, before saturation, the differences in the performance of QHL are negligible relative to the experimental uncertainties in the performance seen in~\fig{badmodel}.  The use of an approximate Hamiltonian model does not substantially degrade the performance of the learning algorithm until the uncertainty in the inference is comparable to the sum of the magnitudes of the neglected couplings.  Such plateaus do not represent a failure of QHL; on the contrary, they point to failures in our modeling of the system and that new physics may be required to understand the system in question.

Conversely, one could also consider the problem of what happens when Bayesian inference is used when there are too many parameters.  It is conceivable in such cases that, rather than outputting the simplest possible model for the Hamiltonian, QHL outputs an unnecessarily complicated model that nonetheless predicts the experimental results with high probability.  Such examples are known to not be typical of Bayesian inference~\cite{MacKay2003Information}.  In fact, Bayesian inference includes Occam's razor by implicitly penalizing unnecessarily complicated models for the data.  We discuss this next.

\subsection{Learning the Best Hamiltonian Model}\label{app:model}
Our results so far have shown that small imperfections do not typically prevent the QHL algorithm from learning the correct Hamiltonian model for a system, but a problem remains: can we use the QHL algorithm to find an accurate and concise model for a unknown quantum system where the form of the Hamiltonian is not even known?  Consider the case that the true model is $\tilde H$, but that we posit the model $H$.  Then, we find that in some sense, the algorithm still learns the ``best'' set of parameters within the set of allowed parameters of the model $H$.  However, since the ``true'' parameters lie outside the set of those allowed by $H$, the distance (as measured by the quadratic loss) of the estimated parameters in $H$ to the true parameters in $\tilde H$ is bounded, as we shown in \fig{badmodel} and \fig{noisecompare}.

This behavior is in fact desirable.  Since modeling physical systems always necessitates some approximation, the optimal estimation procedure ought to find the parameters within the allowed set that is closest to those true parameters outside it.  We can do more, however. In addition to behaving near optimally within each model, our algorithm naturally accommodates \emph{model selection}, whereby it ranks models according to their relative plausibility.  That is, the algorithm simultaneously solves the parameter estimation problem and the meta-problem of finding best model while { minimizing the \emph{effective} number of model parameters used}.  We illustrate this in the case where our hypothetical model $H$ is tested against the true model $\tilde H$.

To this end, we compare the probabilities, given the data, that either $H$ or $\tilde H$ is true: $\Pr(H|D)$ versus $\Pr(\tilde H|D)$.  Using Bayes rule we have
\begin{equation}
\Pr(H|D)= \frac{\Pr(D|H)\Pr(H)}{\Pr(D)}.
\end{equation}
Taking the ratio is then convenient as the normalization factor cancels:
\begin{equation}\label{eq:postodds}
\frac{\Pr(\tilde H|D)}{\Pr(H|D)} = \frac{\Pr(D|\tilde H)}{\Pr(D|H)} \frac{\Pr(\tilde H)}{\Pr(H)},
\end{equation}
which is called the \emph{posterior odds ratio} and forms the basis for comparing models \cite{Jeffreys1939Theory}.  If the posterior odds ratio is larger than 1, the evidence favors $\tilde H$ and vice versa if the value is less than 1.  The last fraction is called the \emph{prior odds}, and the unbiased choice favoring neither model is to set this term equal to 1.  Doing so leaves us with 
\begin{equation}
\frac{\Pr(\tilde H|D)}{\Pr(H|D)} = \frac{\Pr(D|\tilde H)}{\Pr(D|H)},
\end{equation}
which is called the \emph{Bayes factor} \cite{Kass1995Bayes}.

The use of the Bayes factor for model selection is well motivated by other model selection criteria.
The most commonly used model selection technique is the Akaike information criterion (AIC) as it assumes the simple form in $AIC = \max_H \Pr(D|H) - d$, where $d$ is the number of parameters in the model \cite{burnham_model_2002}.  The preferred model is the one with largest value of $AIC$.  Thus, it is clear how models with more parameters are penalized.  The Bayesian approach we advocate above is more generally applicable.  However,  it is less obvious how the Bayes factor include an ``Occam's razor'' to penalize more complex models.  The simplest way to see the effect is to consider the asymptotics of the Bayes factor terms.  Ignoring terms constant in $N$, the asymptotic marginal likelihood is (see, for example \cite{MacKay2003Information})
\begin{equation}
\Pr(D|H) = \max_H \Pr(D|H) - \frac d 2\log N,
\end{equation}
which is the well-known Bayesian information criterion or BIC \cite{burnham_model_2002}.  Noticing the striking similarity to the AIC mentioned above, it is now clear that the Bayesian approach also penalizes extra free parameters.   {This asymptotic form clarifies how additional parameters are penalized; our SMC algorithm approximates the exact (to within numerical accuracy), non--asymptotic distribution}.

For an arbitrary Hamiltonian, $H$, $\Pr(D|H)$ is called the \emph{marginalized likelihood} since we can obtain its value via marginalizing the likelihood function over the model parameters of $H$:
\begin{equation}\label{eq:marginalization}
\Pr(D|H) = \mathbb E_{\vec{x}|H}[\Pr(D|\vec{x};H)].
\end{equation}
This value can be computed online using the likelihood values that are computed in QHL (or more generally a SMC algorithm).  To show this, consider two pieces of data $D=\{d_2,d_1\}$ and
\begin{align}
\Pr(d_2,d_1|H) &= \Pr(d_2|d_1;H)\Pr(d_1|H),\nonumber\\
& = \mathbb E_{\vec{x}|d_1;H}[\Pr(d_2|\vec{x};H)]\mathbb E_{\vec{x}|H}[\Pr(d_1|\vec{x};H)].
\end{align}
These are expectations over the current distribution, which is exactly what the SMC algorithm is designed to efficiently approximate.  One might suspect that being expectations over the likelihood, such calculations would require more costly simulations of the model.  However, note that, under the SMC approximation,
\begin{equation}
\Pr(D|H) = \mathbb E_{\vec{x}|H}[\Pr(D|\vec{x};H)]\approx \sum_{j=1}^{|\{\vec{x}_i\}|} \Pr(D|\vec{x}_j) w_j,
\end{equation}
which is exactly the normalization of the weights after the update rule.  That is, the marginal likelihood is already explicitly calculated as part of the SMC algorithm used in QHL.

A natural way to use the Bayes factor to perform model selection is to simultaneously run two copies of QHL: one using model $H$ and the other using model $\tilde{H}$.  We refer to the model output by using QHL with $H$ as the null model and the model output by QHL with $\tilde{H}$ the alternate model.  The PGH is used to guess experiment using data from the null model at each step of the QHL algorithm by default.  The same experimental parameters are also used in when QHL is applied to the alternate model (even if $H$ and $\tilde{H}$ have fundamentally different forms and/or parameterizations).  The Bayes factor is then computed by taking the expectation values of the likelihoods computed (using data from the quantum simulator) in both cases and dividing the two results.  If this ratio is greater than $1$, then the roles of the null and alternate models are reversed: the alternate model now dictates the choice of experimental parameters in QHL.  These steps are repeated until the uncertainty in the Hamiltonian favored by the posterior odds ratio is sufficiently low.

\begin{figure*}
\begin{minipage}[t]{0.47\linewidth}
\centering
\includegraphics[width=\linewidth]{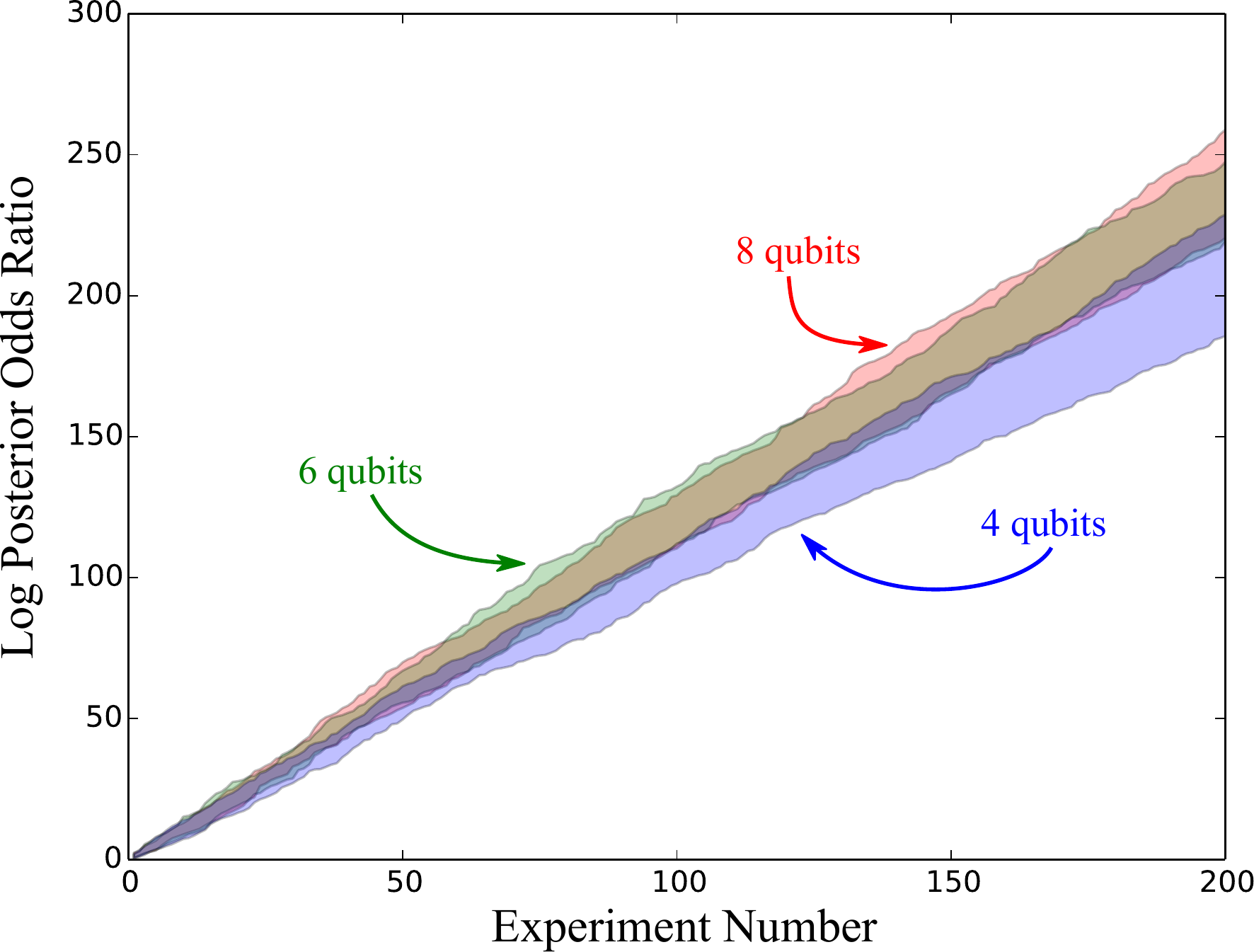}
\caption{The logarithm of the posterior odds ratio of the true model to the reduced model.  Here the reduced model is a Hamiltonian with only  nearest--neighbor couplings on a line while the true Hamiltonian contains small non--nearest neighbor couplings (as described also in \fig{badmodel}).  The bands encompass \emph{all} data.}\label{fig:modelselectcomplete}
\end{minipage}
\begin{minipage}[t]{0.47\linewidth}
\centering
\includegraphics[width=\linewidth]{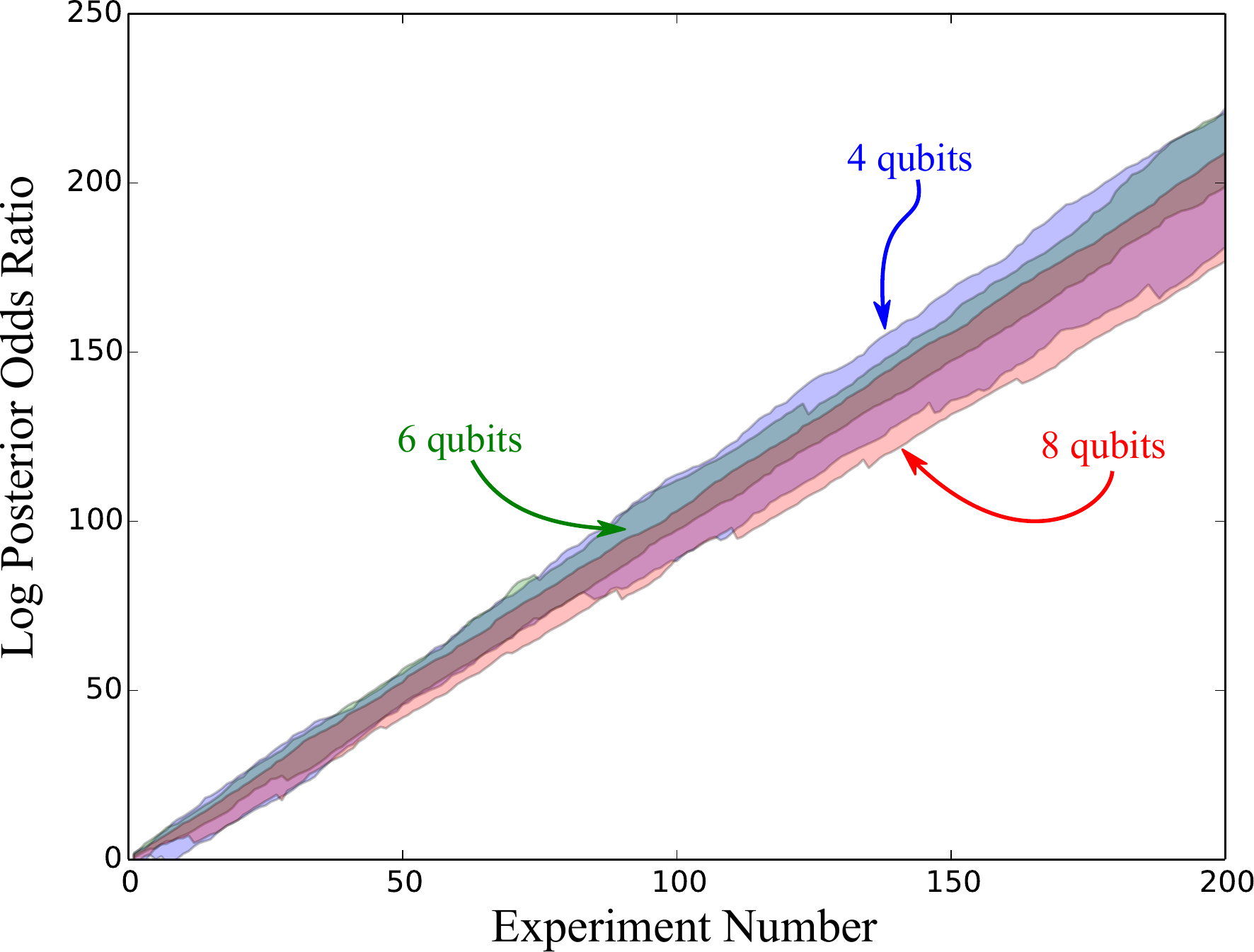}
\caption{The logarithm of the posterior odds ratio for the problem dual to that in \fig{modelselectcomplete}.  Here, the true model contains only nearest--neighbor couplings.  The compared model contains all coupling terms and is hence overfit.  The plot shows that QHL rapidly detects overfitting and urges us to select the reduced model. }\label{fig:modelselectline}
\end{minipage}
\end{figure*}

To illustrate this, consider the example of the previous section (and presented in \fig{badmodel} and \fig{noisecompare}).  The incorrect model learns at an exponential rate but then saturates as the true parameters lie outside the set allowed by the posited model.  Suppose, however, that we use competing models---which could be realized as competing simulators or the same simulator with restricted controls.  To be clear, we take the true Hamiltonian to be an Ising model on the complete graph, $\tilde{H}$ from~\eq{HH'}, and take the model Hamiltonian to be an Ising model on the line, $H$ from~\eq{HH'}.   The initial prior is taken to be uniform over both $a_i$ and the true Hamiltonian is drawn from a similar distribution over $a_i$ and $b_{i,j}$ for each sample.
We then use the Bayesian model selection approach outlined above to decide which model is best.  \fig{modelselectcomplete} shows the logarithm of the posterior odds ratio \eqref{eq:postodds} of the true (non nearest--neighbor Hamiltonian) model to the reduced model (with only nearest--neighbor couplings).  By the 200th measurements, the odds are at least an astounding $10^{120}:1$ against the reduced model suggesting we can also rapidly distinguish good models from bad.  The data for the dual problem---when the true model contains \emph{fewer} parameters---is presented in \fig{modelselectline}.  This corresponds to switching the roles of the true and model Hamiltonians in the previous example.  Again, the algorithm rapidly learns the true model, which in this case is also hedging against overfitting (Occam's razor).

\section{Conclusion}
We show in this paper, both numerically and theoretically, that even imperfect quantum simulators are powerful resources for quantum computation and the characterization of quantum systems.  We show that quantum Hamiltonian learning using interactive likelihood estimation can tolerate substantial amounts of depolarizing before failing to provide useful information about the Hamiltonian.  We also show that realistic errors in the \swapgt~gate do not pose a problem, and that the learning algorithm also can be applied in cases where the model does not commute.  The algorithm is also shown to be robust even in the presence of small unmodelled terms in the actual Hamiltonian; and we see in a numerical example that the algorithm succeeds in finding approximate Hamiltonians that are maximally close to the true Hamiltonian, which has interactions that are not present in the model.  Such cases are particularly intriguing since they can point to failures in the physical models used to describe systems.  The particular way in which the model fails can also be learned by incorporating model selection to distinguish good models from bad.

These results provide a proof of principle that realistic quantum simulators can be used in concert with Bayesian inference to address certain seemingly intractable problems in Hamiltonian estimation, illustrating that quantum resources can be very useful for characterizing and controlling quantum information processors.  This work by no means gives the final answer on Hamiltonian inference, nor does it provide a replacement for strong experimental intuition.  On the contrary, we see that strong understanding of the physics of the system is essential for optimizing the learning rate for the quantum Hamiltonian learning algorithm.  From this perspective, our work raises the possibility of a future where classical machine learning algorithms are employed according to our best knowledge of physics and in tandem with quantum devices, in order to learn properties of unknown systems, certify untrusted quantum devices and perhaps even to discover new physics.

There are a number of natural extensions of this work.  First of all, although the particle guess heuristic often yields very good experiments, it does not necessarily pick ones that are locally optimal.  Locally optimal experiments could be found by minimizing the Bayes' risk using algorithms such as conjugate gradient optimization or differential evolution similar to~\cite{granade_robust_2012,lovett_differential_2013}.  Second, many of the steps in the QHL algorithm could be substantially sped up by using a quantum computer.  A specialized version of the algorithm that incorporates techniques such as amplitude estimation~\cite{brassard2000quantum} and gradient estimation~\cite{jordan2005fast} may show that quantum resources can be leveraged to provide even greater advantages than those considered here.  Finally, although the median quadratic loss tends to behave very well for our algorithm, in relatively rare cases the algorithm can stop learning altogether.  Finding new ways to detect and recover from these errors would be invaluable for reducing the number of times the algorithm must be run  in order to have confidence that the resultant Hamiltonian can actually be trusted.

Our work thus establishes a promising avenue of research in quantum information processing. In particular, our work demonstrates that quantum information processing devices will be useful in the development of further advances in quantum information processing by enabling the use of quantum simulation as a resource. This capability is especially important now,
as the scale of quantum information processing devices grows beyond our classical simulation capacity; hence, the ability to use quantum resources to inexpensively characterize large quantum information processors may prove vital for the development of the next generation of quantum computers and quantum simulators.


\begin{acknowledgements}
We thank Holger Haas for his
  implementation of the pulse finding and cumulant simulation software, and for discussions about physical models. The numerical experiments performed here used SciPy, F2Py and QInfer~\cite{SciPy2001,peterson_f2py:_2009,ferrie_qinfer_2012}.  This work was supported by funding from USARO-DTO, NSERC, CIFAR.  CF was supported by NSF Grant Nos.~PHY-1212445 and ~PHY-1314763 and by the Canadian Government through the NSERC PDF program.
\end{acknowledgements}




\onecolumngrid
\appendix

\section{Parameters Used for Physical Simulations}
\label{app:realistic-sim-params}

We use
the qubit model of \cite{puzzuoli_tractable_2013} to obtain realistic $\swapgt$~gates for superconducting systems.  This model expresses the Hamiltonian as the sum of two component terms given by the single-qubit Hamiltonian
\begin{equation}
  H^{(i)}(t) = \frac{1}{2}[B(t)(1 + \beta_1(t)) + \beta_2(t)] \sigma_z^{(i)}
  + \frac{1}{2}(1 + \alpha(t)) [\cos(\phi(t)) \sigma_x^{(i)} + \sin(\phi(t)) \sigma_y^{(t)}],
\end{equation}
and by the two-qubit interaction Hamiltonian
\begin{equation}
  H^{(ij)} = -\frac{1}{2}C(t)(1 + \gamma(t)) \sigma_z^{(i)} \sigma_z^{(j)}.
\end{equation}
In this model, $A$, $B$, $C$ and $\phi$ are time varying controls, while
$\alpha$, $\beta$ and $\gamma$ are taken to be zero-mean Gaussian processes
with $1/f$ power spectral densities having amplitudes denoted by $\Gamma$
and cutoffs $\Lambda^{(l)}, \Lambda^{(u)}$.

While we mainly consider examples of ``primitive'' gates, consisting
of a single shaped pulse derived from optimal control theory \cite{khaneja_optimal_2005},
we also include a higher-fidelity gate obtained by interlacing with an
XY4 decoupling sequence \cite{gullion_extended_1989} for comparison. This allows us to reason separately
about the impact of the Lindblad generators $L^{(i)} = \frac{1}{2\sqrt{T_1}} (\sigma_x^{(i)} + i \sigma_y^{(i)})$
and the impact of stochastically-varying control fields, given that interlacing
with the XY4 sequence refocuses away much of the stochastic contributions.

We also include an example drawn from the quantum dots model of
\cite{puzzuoli_tractable_2013}. The primary source of noise in this model is the inclusion
of stochasticity in the voltage detuning and Zeeman splitting processes, giving
the single-qubit Hamiltonian
\begin{equation}
    H^{(i)}(t) = \frac12 \frac{A(t) + \alpha(t)}{\sqrt{1 + \exp\left(\frac{B(t)}{B_1} - B_2\right)}} \sigma_x^{(i)}
	       + \frac12 \frac{B(t) - B_0 + \beta(t)}{1 + \exp\left(-\left[\frac{B(t)}{B_1} - B_2\right]\right)} \sigma_z^{(i)},
\end{equation}
where $A$ and $B$ are control parameters for detuning and splitting, respectively,
and where $\alpha$ and $\beta$ are again stochastically-varying noise sources.
We then simulate a
$\swapgt$~gate for the quantum dot model using the parameters for the two qubit gates in \cite{puzzuoli_tractable_2013}.  Apart from
providing an example with a relatively low fidelity swap gate, this example also illustrates that our results are
not predicated on a specific model being used.

In order to generate a range of different gate qualities, and hence demonstrate
the effectiveness of our algorithm in a variety of different physically realistic
scenarios, we increase the noise from those gates used in \cite{puzzuoli_tractable_2013}.
In particular, we shorten the relaxation time $T_1$, and increase the $1/f$-noise amplitudes $\Gamma$.
The former causes the dissipative process acting on our system to become stronger,
while the latter increases the stochasticity of the control fields.
Varying noise parameters in this way, we show gates with fidelities ranging from $F \approx 0.9$ to nearly
ideal. In \tab{realistic-sim-noiseparams}, we list the noise parameters used in the numerical experiments.

\begin{table}[t!]
    \centering
    \caption{\label{tab:realistic-sim-noiseparams} Noise parameters used for superconducting
    gates.}
    \begin{tabular}{lrrrrrrrrl}
      \hline
                                & \multicolumn{1}{c}{XY4}   & \multicolumn{4}{c}{Primitive} & \\
      \hline
      Fidelity w/ $\swapgt$     & $0.998$                   & $0.996$                   & $0.954$                   & $0.906$ \\
      Discretization timestep   & $2.5\times10^{-10}$       & $1\times10^{-10}$         & $1\times10^{-10}$         & $1\times10^{-10}$         & s  & \\
      \hline
      $T_1$                     & $10^{-4}$                 & $10^{-5}$                 & $10^{-5}$                 & $10^{-5}$                 & s   \\
      $\Gamma_{\alpha}$         & $3 \times 10^4$           & $10^4$                    & $10^6$                    & $10^6$                    & Hz  \\
      $\Gamma_{\beta_1}$        & $3 \times 10^4$           & $0$                       & $10^6$                    & $10^6$                    & Hz  \\
      $\Gamma_{\beta_2}$        & $10^6 / 2 \pi$            & $10^4$                    & $10^6$                    & $1.5 \times 10^6$         & Hz  \\
      $\Lambda_{\alpha}^{(l)}$  & $1 / 2 \pi$               & $1 / 2 \pi$               & $1 / 2 \pi$               & $1 / 2 \pi$               & Hz  \\
      $\Lambda_{\alpha}^{(u)}$  & $10^9$                    & $10^9$                    & $10^9$                    & $10^9$                    & Hz  \\
      $\Lambda_{\beta_1}^{(l)}$ & $1 /2 \pi$                & $1 / 2 \pi$               & $1 / 2 \pi$               & $1 / 2 \pi$               & Hz  \\
      $\Lambda_{\beta_1}^{(u)}$ & $10^9$                    & $10^9$                    & $10^9$                    & $10^9$                    & Hz  \\
      $\Lambda_{\beta_2}^{(l)}$ & $1 / 2 \pi$               & $1 / 2 \pi$               & $1 / 2 \pi$               & $1 / 2 \pi$               & Hz  \\
      $\Lambda_{\beta_2}^{(u)}$ & $10^9$                    & $10^9$                    & $10^9$                    & $10^9$                    & Hz  \\
      $\Gamma_{\gamma}$         & $1.2 \times 10^3 / 2 \pi$ & $1.2 \times 10^3 / 2 \pi$ & $1.2 \times 10^5 / 2 \pi$ & $1.2 \times 10^5 / 2 \pi$ & Hz  \\
      $\Lambda_{\gamma}^{(l)}$  & $1 / 2 \pi$               & $1 / 2 \pi$               & $1 / 2 \pi$               & $1 / 2 \pi$               & Hz  \\
      $\Lambda_{\gamma}^{(u)}$  & $10^9$                    & $10^9$                    & $10^9$                    & $10^9$                    & Hz  \\
      \hline                          
    \end{tabular}
\end{table}


\end{document}